\documentclass[12pt,a4paper]{amsart}
\usepackage{amsmath,amssymb,a4wide,amsthm,ifthen,hyperref,xcolor,enumerate,enumitem}
\usepackage[utf8]{inputenc}

\usepackage{graphicx}
\usepackage{url}
\usepackage{ulem}

\usepackage{epstopdf}
\usepackage[section]{placeins}
\usepackage{natbib}
\usepackage{flafter} 
\usepackage{multirow}

\begin{document}

\title[A novel regularized approach for functional data clustering]{A novel regularized approach for functional data clustering: An application
  to milking kinetics in dairy goats}


\author{C. Denis}
\address{UMR MIA-Paris, AgroParisTech, INRA, Universit\'e 
Paris-Saclay, 75005, Paris, France and LAMA - Université Paris-Est - Marne-la-Vallée, 77420 Champs-sur-Marne, France}
\email{christophe.denis@u-pem.fr}
\author{E. Lebarbier}
\address{UMR MIA-Paris, AgroParisTech, INRA, Universit\'e 
Paris-Saclay, 75005, Paris, France}
\email{emilie.lebarbier@agroparistech.fr}
\author{C. L\'evy-Leduc}
\address{UMR MIA-Paris, AgroParisTech, INRA, Universit\'e 
Paris-Saclay, 75005, Paris, France}
\email{celine.levy-leduc@agroparistech.fr}
\author{O. Martin}
\address{UMR Modélisation Systémique Appliquée aux Ruminants, INRA, AgroParisTech, Université Paris-Saclay, 75005, Paris, France}
\email{olivier.martin@agroparistech.fr}
\author{L. Sansonnet}
\address{UMR MIA-Paris, AgroParisTech, INRA, Universit\'e 
Paris-Saclay, 75005, Paris, France}
\email{laure.sansonnet@agroparistech.fr}

\maketitle

\begin{abstract}
  Motivated by an application to the clustering of milking kinetics of dairy goats, we propose in this paper a novel approach for functional data clustering.
This issue is of growing interest in precision livestock farming that has been largely based on the development of data acquisition automation and on the development of
 interpretative tools to capitalize on high-throughput raw data and to generate benchmarks for phenotypic traits. The method that we propose in this paper falls in this context.
  Our methodology relies on a piecewise linear estimation of curves based on a novel
  regularized change-point estimation method
and on the $k$-means algorithm applied to a vector of coefficients summarizing the curves. The statistical performance of our method is assessed through
numerical experiments and is thoroughly compared with existing ones.
Our technique is finally applied to milk emission kinetics data with the aim of a better characterization of inter-animal variability and toward a better
understanding of the lactation process.
\end{abstract}

\section{Introduction}

Precision livestock farming is a blooming field grounded in the development of sensors providing high throughput data and thus potentially increasing access to valuable information on biological processes. Therefore, developing methods for data analysis and interpretation has become a challenging issue in animal science. Economic performance of dairy goat farming systems is primarily based on milk production and a large amount of farmers working time is spent milking animals, see \cite{marnet:2005}. Moreover, with the increasing size of goat herds and the rapid growth of the dairy goat industry, more in-depth information on individual milking performance is necessary. In this context, a better understanding of the variability in milk flow kinetics could for instance help refining selection criteria for breeding programs, simplifying milking workload or controlling udder health. Milk emission kinetics recorded during milking of dairy goats are classically described and classified through synthetic parameters such as milking time, maximum and average milk flow rates, time to reach 500 g/min milk flow, see \cite{romero:panzalis:ruegg:2017}. In this paper, we explore the possibility of considering milk emission kinetics as a whole function, opening new perspectives to study inter-animal variability.

From a statistical point of view, this issue belongs to the general field of functional data analysis, see \cite{Ramsay:silverman:2005} for a survey on this subject.
In the specific functional data clustering framework, 
several approaches have been proposed by \cite{abraham:cornillon:2003}, \cite{jacques:2013} and \cite{bouveyron2015} among others. 
For a review on this subject, we refer the reader to \cite{jacques:preda:survey:2014} and the references therein.
This kind of approaches was extended to deal with multivariate functional
data by \cite{jacques:preda:2014} who proposed the first model-based clustering algorithm in this multivariate context and more recently by \cite{schmutz:jacques:bouveyron:2018}.

\begin{figure}
\includegraphics[scale=0.26]{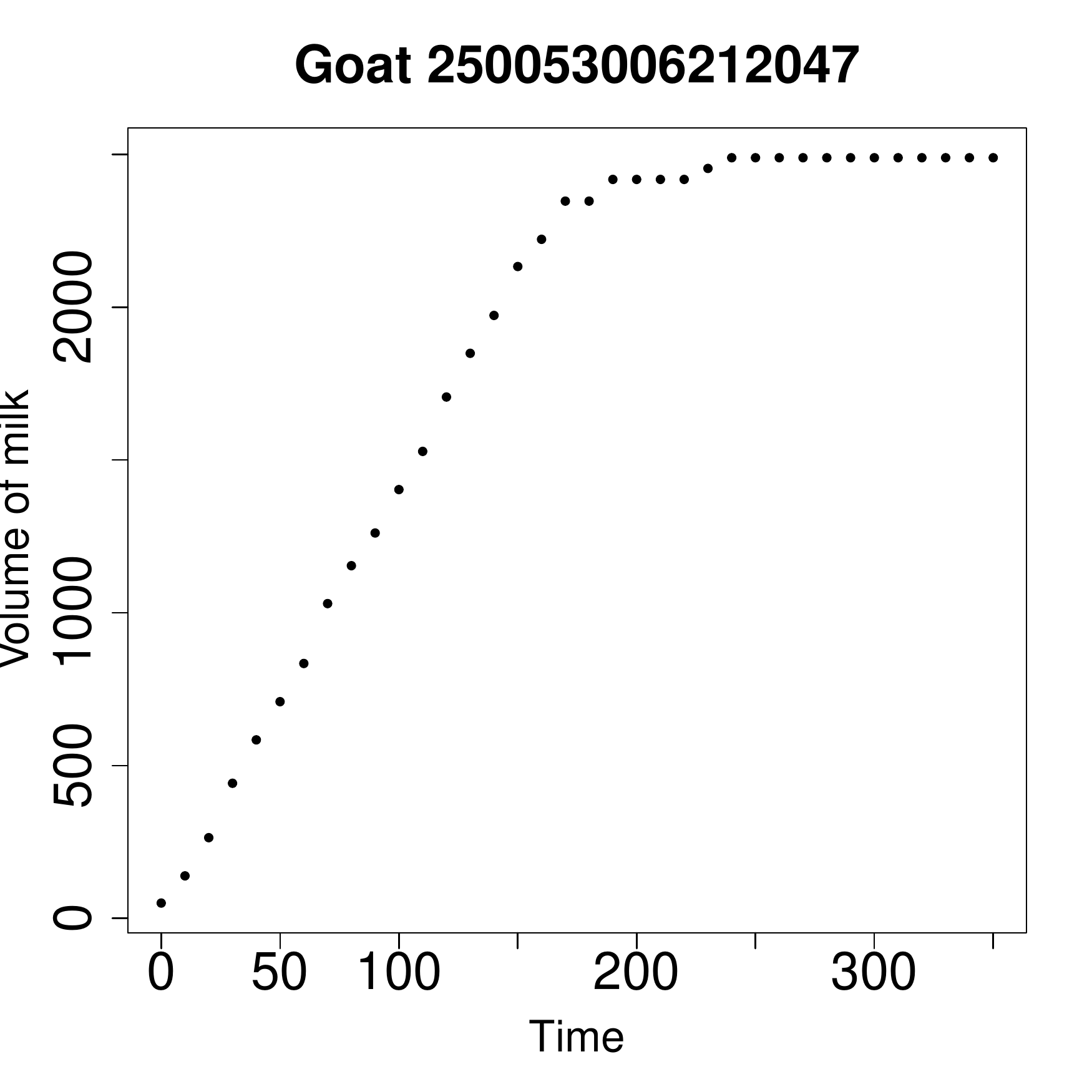}
\includegraphics[scale=0.26]{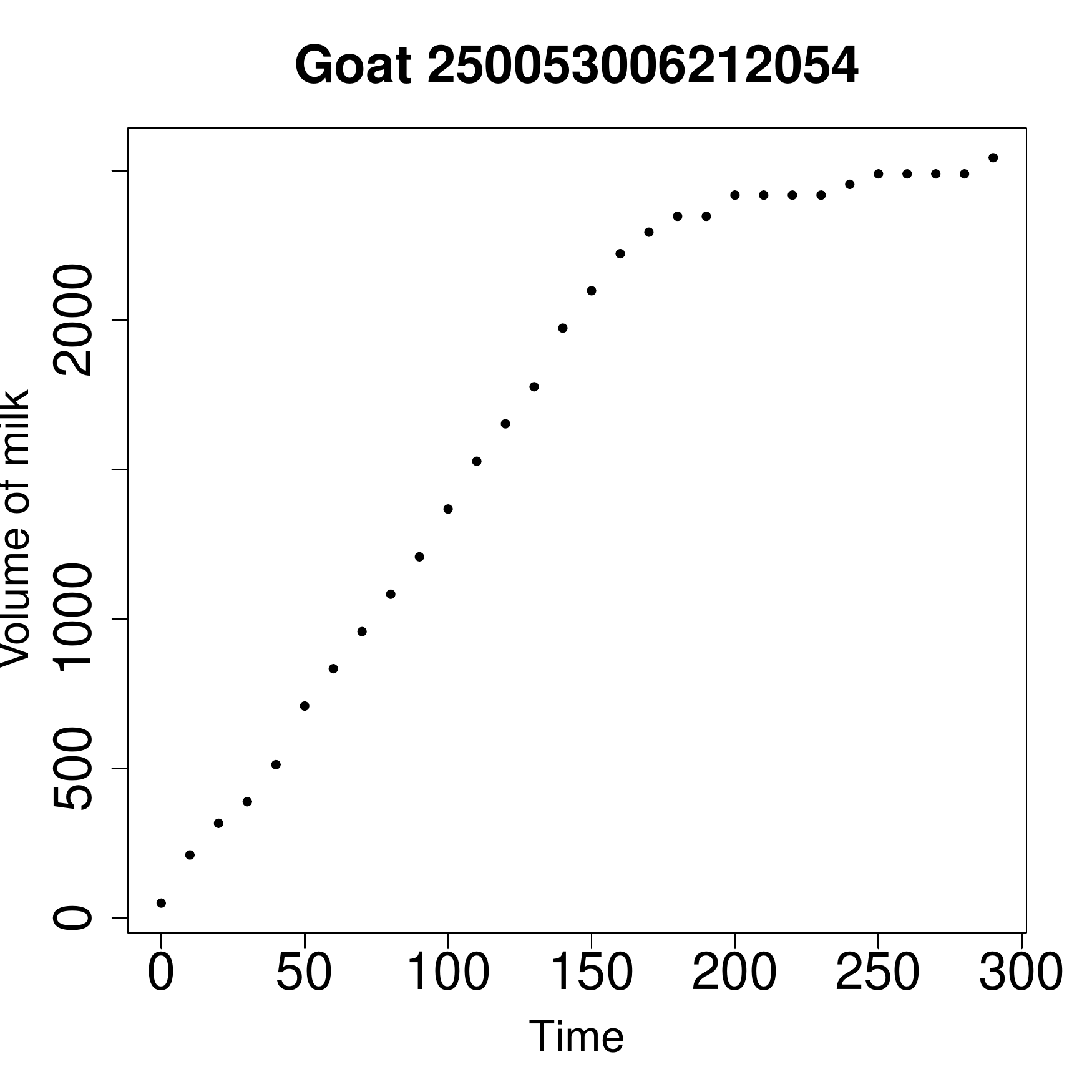}
\includegraphics[scale=0.26]{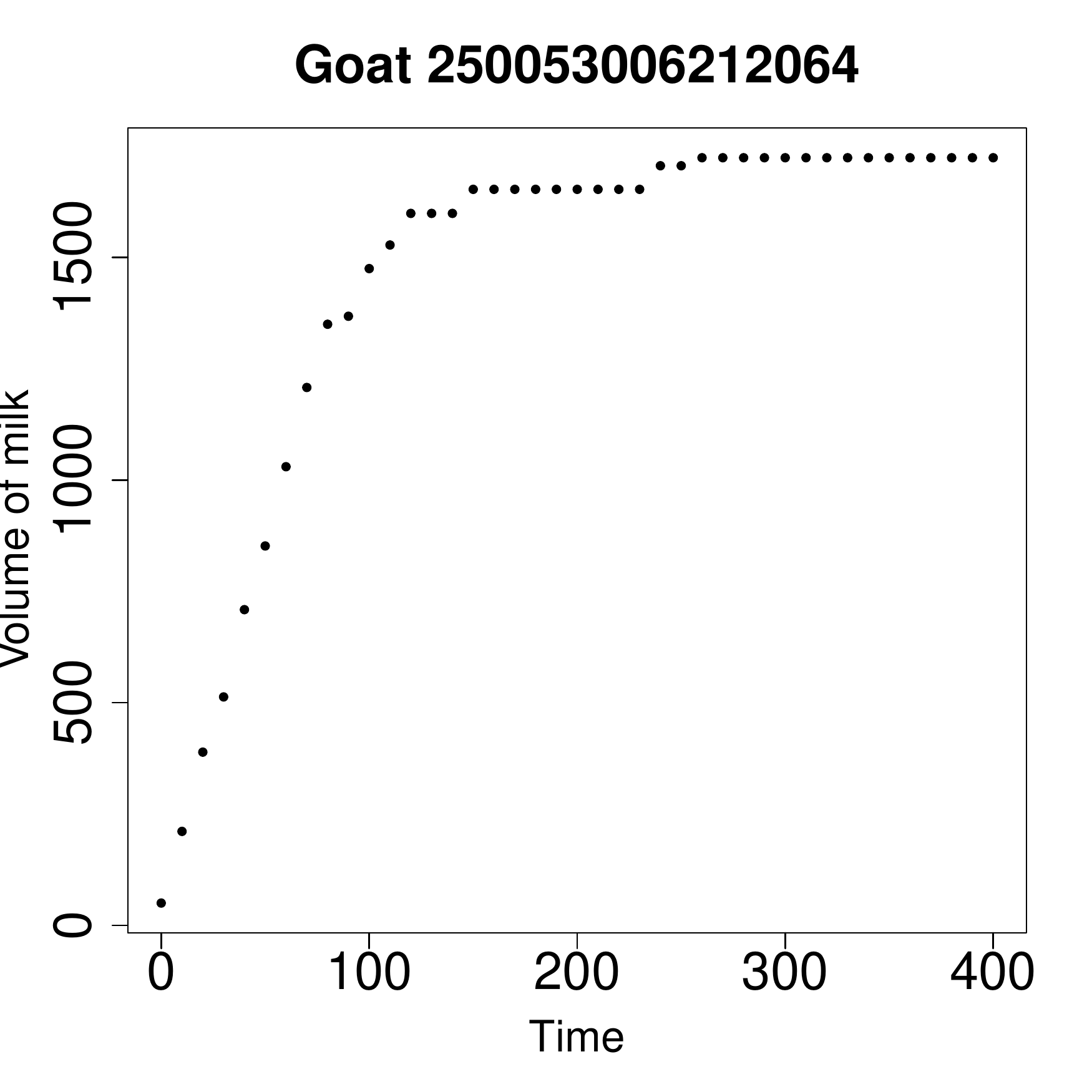}\\
\includegraphics[scale=0.26]{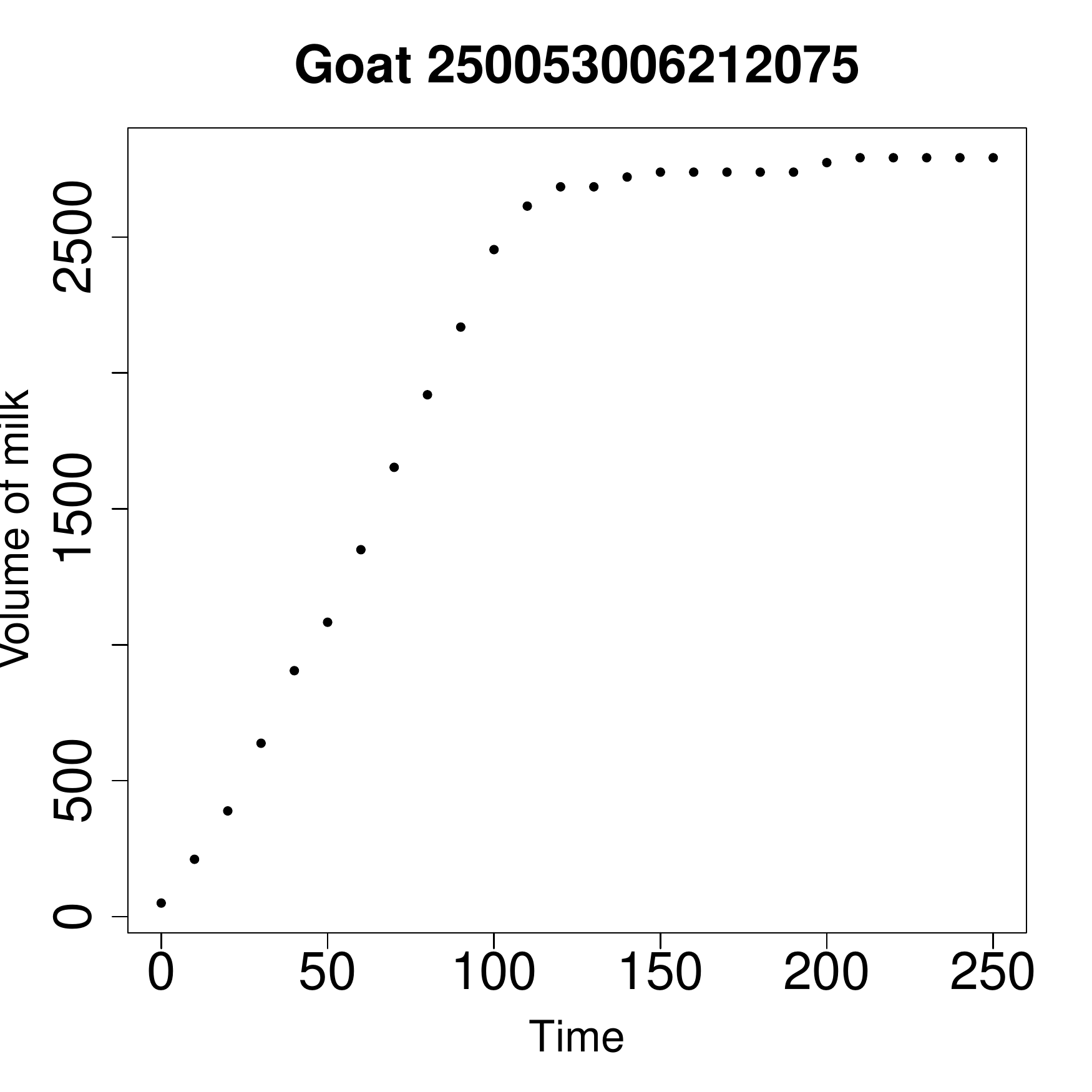}
\includegraphics[scale=0.26]{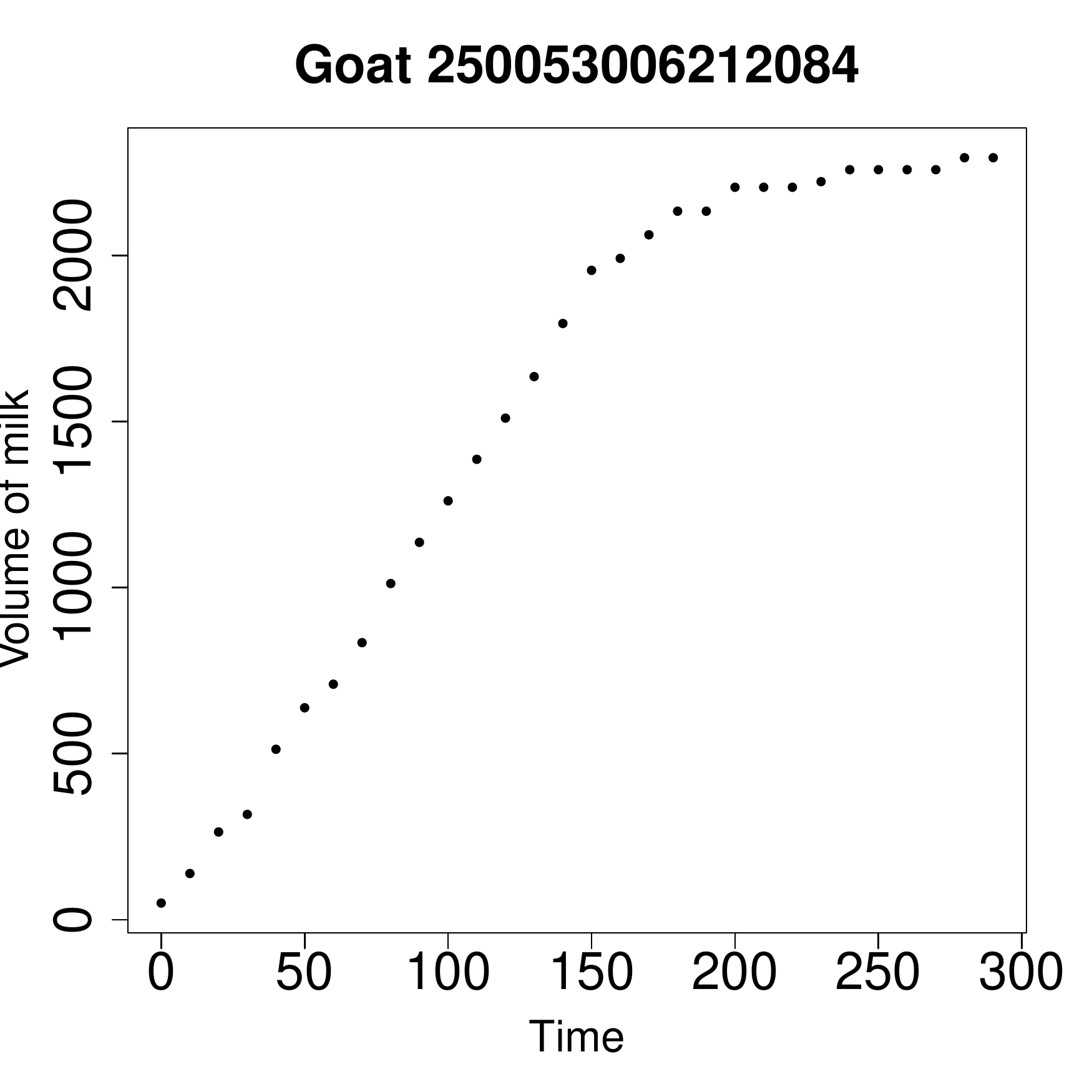}
\includegraphics[scale=0.26]{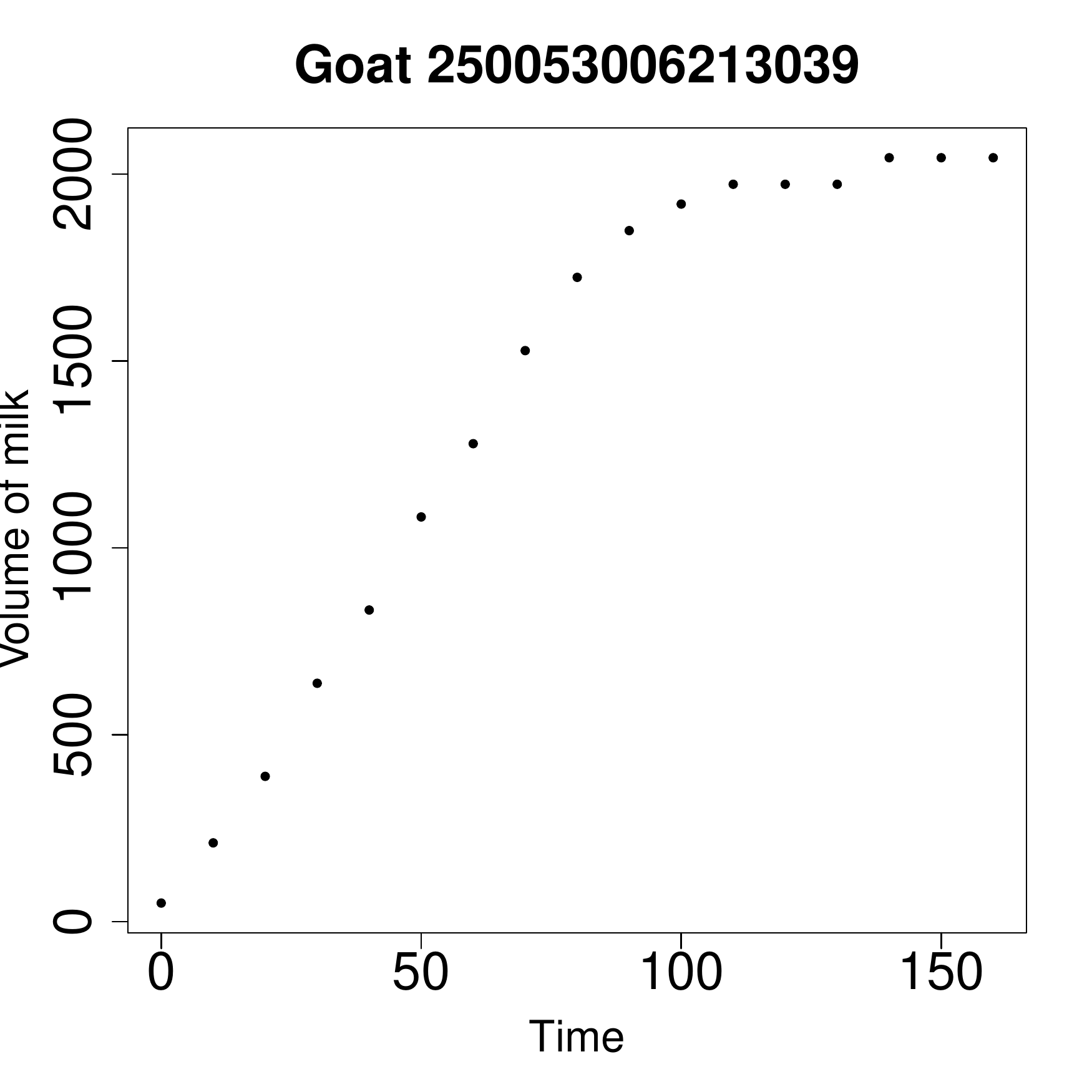}
\caption{Some examples of milking kinetics of goats.\label{fig1}}
\end{figure}

To deal with the functional clustering of the milking kinetics of goats, some specific features have to be taken into account, see
Figure \ref{fig1} for some examples of such kinetics.
We can see from this figure that these curves are
nondecreasing and can be split into two parts, namely an increasing linear part and an almost constant one. Inspired by \cite{abraham:cornillon:2003}, we  propose
in this paper a dimension reduction approach based on a continuous piecewise linear function fit to each curve which boils down to a change-point detection issue
which will be crucial in our method.

The problem of detecting change-points in the mean of a signal is largely addressed in the literature. In particular, it is now well known that in (penalized-) maximum likelihood frameworks the Dynamic Programming (DP) algorithm (\cite{Bellman:1961,Auger89}) and its recent pruned versions \cite{killick_pelt,G15, Maidstone2016} are the only algorithms that retrieve the exact solution very quickly. However, DP can only be used  if the contrast to be optimized is additive with respect to the segments, see for example \cite{BP03,PRL05,lavielle:2005}. When detecting changes in the slope with a continuity condition, the segments will 
unavoidably be linked and therefore the additivity condition is not satisfied.
This partly explained that this change-point detection problem has not been thoroughly investigated in the literature compared to the simplest detection in the mean problem. Recently, \cite{fearnhead2019detecting} proposed to extend the PELT algorithm \cite{killick_pelt} to this problem. Their idea is to include the penalty 
in the DP algorithm with a pruning strategy. The penalty they proposed is proportional to the number of change-points up to a penalty constant. However, this penalty constant needs to be chosen in 
advance, which is not easy in practical situations. 

In this paper, we first propose a novel  change-point estimation in the slope method combining the trend filtering proposed by \cite{tibshirani:2014} with
a (penalized-) maximum likelihood  approach which is useful for removing the spurious change-points that may have been proposed by trend filtering.  
These change-points estimators are then used for devising a new dimension reduction approach: Each curve is summarized by a vector containing the coefficients of its projection 
onto an order 2 $B$-spline basis having for knots the obtained change-points and also the change-point locations. Including the change-points both in the features characterizing the curves 
and in the $B$-spline knots is the main novelty compared to classical approaches reviewed in \cite{jacques:preda:survey:2014}.

The paper is organized as follows. The methodology that we propose is decribed in Section \ref{sec:methodology}. 
The performance of our approach is investigated in Section \ref{sec:numExp} through numerical
experiments. Finally, in Section \ref{sec:appli}, we apply our method to the data that motivated this study.


\section{Methodology}\label{sec:methodology}

In this section, we describe our novel functional data clustering approach which consists of two steps which can be summarized as follows:

\begin{itemize}
\item \textsf{First step:} Piecewise linear estimation of the curves using a novel change-point estimation method based on the trend filtering approach and $B$-splines.
\item \textsf{Second step:} Applying the $k$-means algorithm  to a vector of coefficients summarizing the curves obtained in the first step.
\end{itemize}

These two steps are further described hereafter.

\subsection{First step: Piecewise linear estimation of the curves based on a change-point estimation method}

In the following, we assume that the observations of a given curve $\mathbf{Y}=(Y_1,\ldots,Y_n)$ correspond to a noisy function evaluated at the input points
$\mathbf{x}=(x_1,\dots,x_n)$.
In this step, we aim at estimating each curve by a piecewise linear function using a two-stage approach described below.

\subsubsection{First stage: Trend filtering for change-point estimation}
We use the trend filtering approach proposed by \cite{tibshirani:2014} which consists in fitting to the observations $\mathbf{Y}$
the vector $\widehat{\boldsymbol{\beta}}=(\widehat{\beta}_1,\dots,\widehat{\beta}_n)$ using a regularized method. More precisely, we use
$$
\widehat{\boldsymbol{\beta}}(\lambda)=\textrm{Argmin}_{\boldsymbol{\beta}\in\mathbb{R}^n}\left\{\|\mathbf{Y}-\boldsymbol{\beta}\|_2^2+\lambda \|D^{(2)}\boldsymbol{\beta}\|_1\right\},
$$
where $\|y\|_2^2=\sum_{i=1}^n y_i^2$, $\|y\|_1=\sum_{i=1}^n |y_i|$, for $y=(y_1,\dots,y_n)$, 
$\lambda$ is a positive constant which has to be tuned and $D^{(2)}$ is the discrete difference operator of order 2 defined by
$$
D^{(2)}=
\left(
\begin{matrix}
1 & -2 & 1  & 0 & 0 &\cdots  & 0\\
0 &  1 & -2 & 1 & 0 & \cdots & 0\\
\vdots &  &\ddots  & \ddots & \ddots &  & \vdots\\
\end{matrix}
\right).
$$

The final estimator of $\boldsymbol{\beta}$ is $\widehat{\boldsymbol{\beta}}(\widehat{\lambda})$ where $\widehat{\lambda}$ has to be properly chosen. 
Usually, this parameter is chosen using resampling approaches such as cross-validation or stability selection, see \cite{meinshausen:buhlmann:2010}. 
From $\widehat{\boldsymbol{\beta}}(\widehat{\lambda})$, we define a set of potential change-point indices as the coordinates where the vector $D^{(2)}\widehat{\boldsymbol{\beta}}(\widehat{\lambda})$
is not equal to zero. However, in change-point 
estimation frameworks, the performance of such methods may be altered since some change-points may be omitted 
by subsampling. 
Moreover, it is well known that such regularization approaches lead to over-segmentation phenomena. 
Usually, in this case, a DP algorithm is then used on the set of potential change-points obtained with the latter strategy in order to remove the irrelevant ones, 
see for instance \cite{harchaoui:levy:nips} and \cite{harchaoui:levy:jasa:2010}. 

We propose following this strategy: In order to avoid the use of a resampling method, we choose a 
small enough $\lambda$ in order to obtain a large enough set of potential change-points. More precisely, we set a maximal number of change-points denoted 
$K_{\textrm{max}}$ and choose $\lambda$ such that among the $\lambda$'s leading to $K_{\textrm{max}}$ 
change-points, $\widehat{\lambda}$ is the one minimizing $\|\mathbf{Y}-\widehat{\boldsymbol{\beta}}(\lambda)\|_2^2$.


\textcolor{black}{Let $(\widehat{n}_1,\dots,\widehat{n}_{K_{\textrm{max}}})$ the resulting change-point indices and the associated change-point positions
$(\widehat{t}_1,\dots,\widehat{t}_{K_{\textrm{max}}})=(x_{\widehat{n}_1},\dots,x_{\widehat{n}_{K_{\textrm{max}}}})$. For each $K$ in $\{1,\dots,K_{\textrm{max}}\}$, we use the DP algorithm to retrieve 
the $K$ most relevant change-point indices
among $\widehat{n}_1,\dots,\widehat{n}_{K_{\textrm{max}}}$. DP is thus applied to $Y_{\widehat{n}_1},\dots,Y_{\widehat{n}_{K_{\textrm{max}}}}$ instead of $Y_1,\dots,Y_n$.
Note that a slight modification of the algorithm is considered to make the piecewise linear fit to data continuous.
The optimal number of change-points $\widehat{K}$ is then chosen by using the criterion proposed by \cite{lavielle:2005}.}

\subsubsection{Second stage:  Projection onto the $B$-spline basis having as knots the obtained change-points}



Each curve will then be summarized by a few coefficients corresponding to the coefficients of its projection onto
the $B$-spline basis $(B_{i,2})_{1\leq i\leq\widehat{K}+2}$ defined as follows, see \cite[p. 206]{hastie:tibshirani:friedman:2009} for a review on the subject.
Let $\widehat{t}_0=x_1$ and $\widehat{t}_{\widehat{K}+1}=x_n$. Let us also define the augmented knot sequence $\tau$ such that:
$$
\tau_1=\tau_2=\widehat{t}_0=x_1,
$$
$$
\tau_{j+2}=\widehat{t}_j,\; j=1,\dots,\widehat{K},
$$
$$
\tau_{\widehat{K}+3}=\tau_{\widehat{K}+4}=\widehat{t}_{\widehat{K}+1}=x_n,
$$
namely,
$$
(\tau_1,\dots,\tau_{\widehat{K}+4})=(x_1,x_1,\widehat{t}_1,\cdots,\widehat{t}_{\widehat{K}},x_n,x_n).
$$
The $i$th $B$-spline function $B_{i,2}$ having $\tau$ for knot sequence
satisfies:
$$
B_{i,2}(u)=\frac{u-\tau_i}{\tau_{i+1}-\tau_i}B_{i,1}(u)+\frac{\tau_{i+2}-u}{\tau_{i+2}-\tau_{i+1}} B_{i+1,1}(u),
$$
where
$$
B_{i,1}(u)=
\begin{cases}
  1, \textrm{ if } \tau_i\leq u <\tau_{i+1}\\
  0,\textrm{ otherwise}
\end{cases}
$$
with $i\in\{1,\dots,\widehat{K}+2\}$.
Thus, each curve is estimated by $\widehat{f}$ defined by:
\begin{equation}\label{eq:f_Bspline}
\widehat{f}(u)=\sum_{i=1}^{\widehat{K}+2}\widehat{\theta}_i B_{i,2}(u),
\end{equation}
where the $\hat{\theta}_i$'s are obtained using a least-square criterion.
Hence, the coefficients summarizing each curve is: 
\begin{equation}\label{eq:summary}
  (\widehat{\theta}_1,\dots,\widehat{\theta}_{\widehat{K}+2},\widehat{t}_1,\dots,\widehat{t}_{\widehat{K}}).
\end{equation}

\subsection{Second step: Clustering using the $k$-means algorithm}\label{subsec:second_step}

In order to obtain a clustering of the curves (milking kinetics), we use the $k$-means algorithm of \cite{hartigan1979algorithm}
on the scaled summarized coefficients (\ref{eq:summary}) obtained in the previous step. It has to be noticed that the number of change-points $\widehat{K}$
may change from one curve to the other. Thus, we consider summarized coefficients of length $\widehat{K}_M$ corresponding to the largest value
of $\widehat{K}$. For kinetics having a number of change-points smaller than $\widehat{K}_M$, we replace the missing $\widehat{t}_k$ and the
missing coefficients by 0. Our goal is indeed to propose a strategy which is able to distinguish the curves both thanks to the change-point positions and/or 
the coefficient values.

The number $k$ of clusters  is chosen by using the strategy proposed by \cite{nbclust:2014} which consists in using the majority rule that is taking for $k$
the value chosen by the largest number of criteria among 30 indices such as: CH index, Duda index, Pseudot2 index, C index, Hartigan index, ...
Further details on these indices can be found in \cite{nbclust:2014}.
Here, we focused on the four following indices: KL index, Hartigan index, SDindex, Ptbiserial index.


\section{Numerical experiments}\label{sec:numExp}

In this section, we investigate the statistical performance of our procedure.
The simulation scheme that we used for this investigation is described in Section~\ref{subsec:simulScheme}.
We also propose in Section~\ref{subsec:numericalPerf} to benchmark our procedure with existing approaches
and to assess our change-point estimation approach in Section \ref{sec:assessment}.

\subsection{Simulation scheme}
\label{subsec:simulScheme}

In order to be as close as possible to the data coming from our motivating application, 
we consider two  different models for generating the data 
that we will refer to as $\texttt{Model 1}$ and $\texttt{Model 2}$ in the following. For each model, the complete observed data is $(\mathbf{Y},Z)$,
where $\mathbf{Y}$ is in $\mathbb{R}^n$ and corresponds to the observations of an underlying function, which we will specify hereafter, at the input points 
$\mathbf{x}=(x_i)_{1\leq i\leq n}=(10(i-1))_{1\leq i\leq n}$ with $n=51$.  
$Z$ denotes the label of $\mathbf{Y}$ which takes its value in $\mathcal{Z}=\{1,2,3,4\}$.
Moreover, for each $z \in \mathcal{Z}$, the associated cluster $\mathcal{C}_z$ is characterized by a number of change-points $K_z$, a vector of change-points $t^z$, and a vector of parameters 
$\theta^z \in \mathbb{R}^{K_z + 1}$. Hence,
each model is defined by a set of parameters $\{K_z, t^z, \theta^z : \; z \in \mathcal{Z}\}$. The values of the parameters associated to each model are reported in Tables~\ref{table:tableParameters1}
and \ref{table:tableParameters2}. 
Note that for each model, the clusters are distinguishable by both the change points and the parameters.

\begin{table}
\caption{Set of parameters for $\texttt{Model 1}$.\label{table:tableParameters1}}
\begin{center}
\begin{tabular}{|c|c|c|c|}
\hline
\multicolumn{4}{|c|}{$\texttt{Model 1}$} \\
\hline
$z$ & $K_z$ & $t^z$ & $\theta^z$ \\
\hline
1 & 2 & $(150,250)$ & $(1600,1900,2000)$ \\
\hline
2 & 2 & $(150,300)$ & $(1400,1800,2200)$ \\
\hline
3 & 4 & $(100,200,300,400)$ & $(300,1500,1700,2000,2200)$ \\
\hline
4 & 3 & $(50,150,300)$ & $(200,1300,1800,2100)$ \\
\hline
\end{tabular}
\end{center}
\end{table}

\begin{table}
\caption{Set of parameters for $\texttt{Model 2}$.\label{table:tableParameters2}}
\begin{center}
\begin{tabular}{|c|c|c|c|c|}
\hline
\multicolumn{4}{|c|}{$\texttt{Model 2}$} \\
\hline
$z$ & $K_z$ & $t^z$ & $\theta^z$ \\
\hline
1 & 2 & $(150,250)$ & $(1600,1900,2000)$ \\
\hline
2 & 2 & $(150,300)$ & $(1400,1800,2200)$ \\
\hline
3 & 4 & $(100,200,300,400)$ & $(300,1500,1700,2000,2200)$ \\
\hline
4 & 3 & $(150,250,300)$ & $(200,700,1000,1600)$ \\
\hline
\end{tabular}
\end{center}
\end{table}

For each model, the vector $(\mathbf{Y},Z)$ is simulated according to the following procedure:
\begin{enumerate}
\item[(a)]  The label $Z$ is drawn from a uniform distribution on $\mathcal{Z}$;
\item[(b)]  We generate $\tilde{t}^Z = t^Z + \mathcal{U}$, such that $\mathcal{U}=(U, \ldots, U)$, where $U$ is a uniformly distributed random variable on $\{-30,-20,10,0,10,20,30\}$;   
\item[(c)] We generate $\tilde{\theta}^Z = \theta^Z + \mathcal{V}$, such that $\mathcal{V} = (V, \ldots, V)$, where $V$ is a uniformly distributed random variable on $[-200,200]$;
\item[(d)] Then, we consider the sequences $(\tilde{t}^Z_0, \ldots, \tilde{t}^Z_{K_Z + 1}) = (0, \tilde{t}^Z, 500)$, $(\tilde{\theta}^Z_{0},\ldots, \tilde{\theta}^{Z}_{K_Z + 1}) = (0,\tilde{\theta}^Z)$, and define for $x \in [\tilde{t}^Z_j, \tilde{t}^Z_{j+1}]$, and $j \in \{0,\ldots, K_Z\}$
\begin{equation}\label{eq:ftilde}
f_{\tilde{t}^Z, \tilde{\theta}^Z}(x) = (\tilde{\theta}^{Z}_{j+1}-\tilde{\theta}^{Z}_{j}) \dfrac{x-\tilde{t}^Z_j}{\tilde{t}^Z_{j+1}-\tilde{t}^Z_j} + \tilde{\theta}^{Z}_{j};
 \end{equation} 
 \item[(e)] Finally, we define $\mathbf{Y}$ such that, for $i \in \{1, \ldots, n\}$, 
\begin{equation}
\label{eq:eqSimY}
 Y_i = f_{\tilde{t}^Z, \tilde{\theta}^Z}(x_i) + \varepsilon_i,
 \end{equation}
 where the $\varepsilon_i$'s are i.i.d $\mathcal{N}(0,\sigma^2)$ random variables with $\sigma\in\{1,5\}$.
\end{enumerate}
Note that the function $f$ defined in (\ref{eq:ftilde}) can be seen as another way of writing (\ref{eq:f_Bspline}).

\begin{figure}
\begin{center}
\begin{tabular}{cc}
\includegraphics[width = 7.5cm, scale=1]{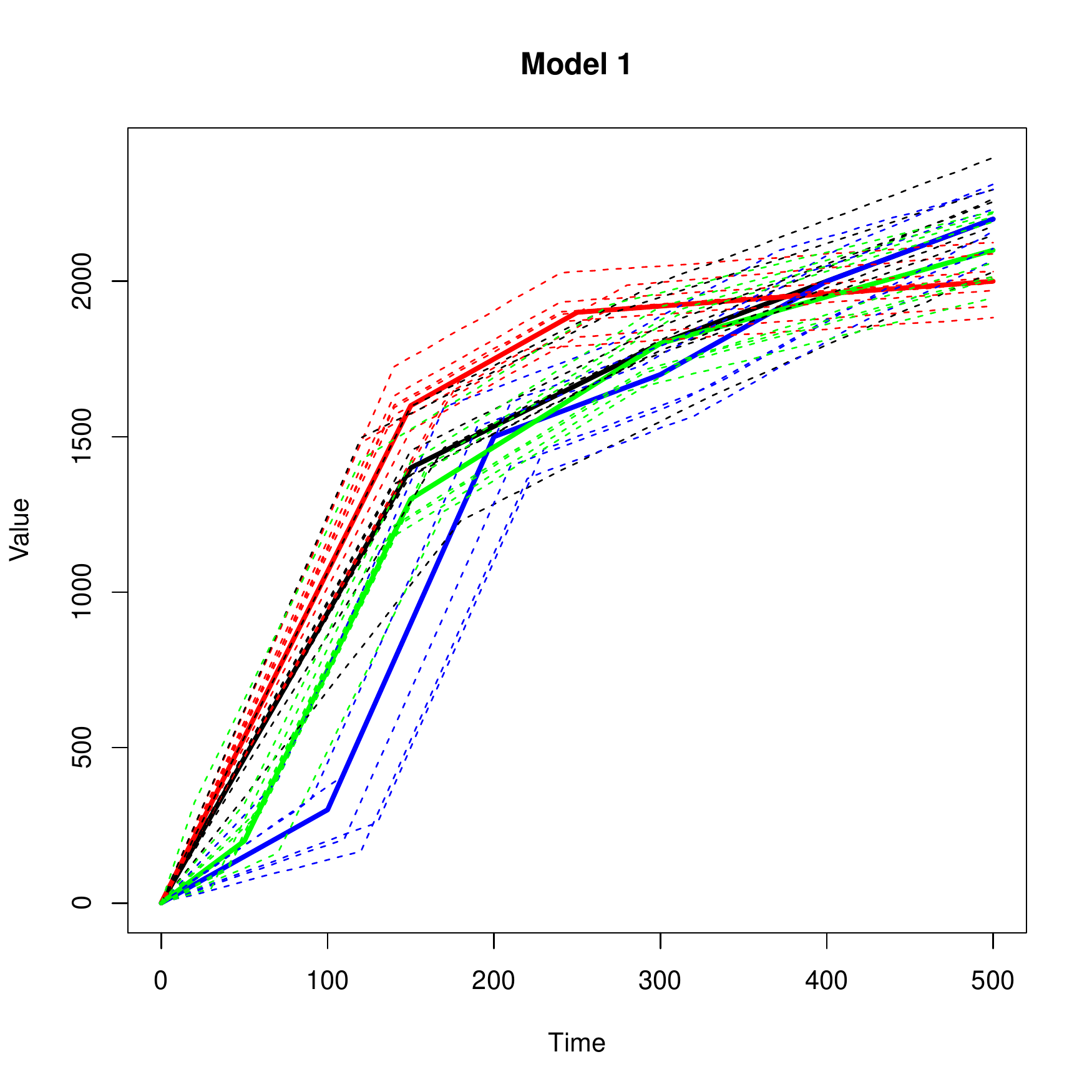} & \includegraphics[width = 7.5cm, scale=0.1]{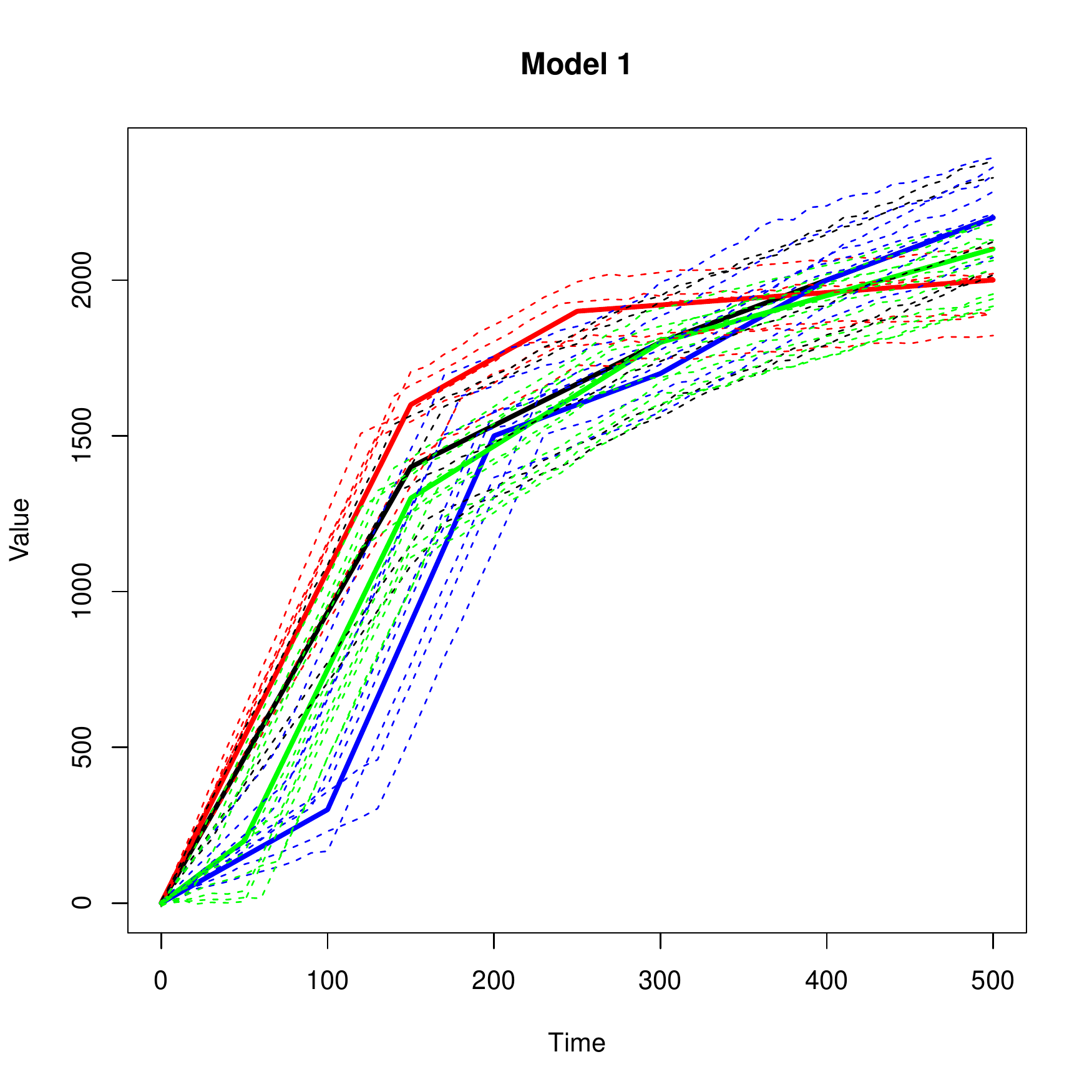}\\
\includegraphics[width = 7.5cm, scale=1]{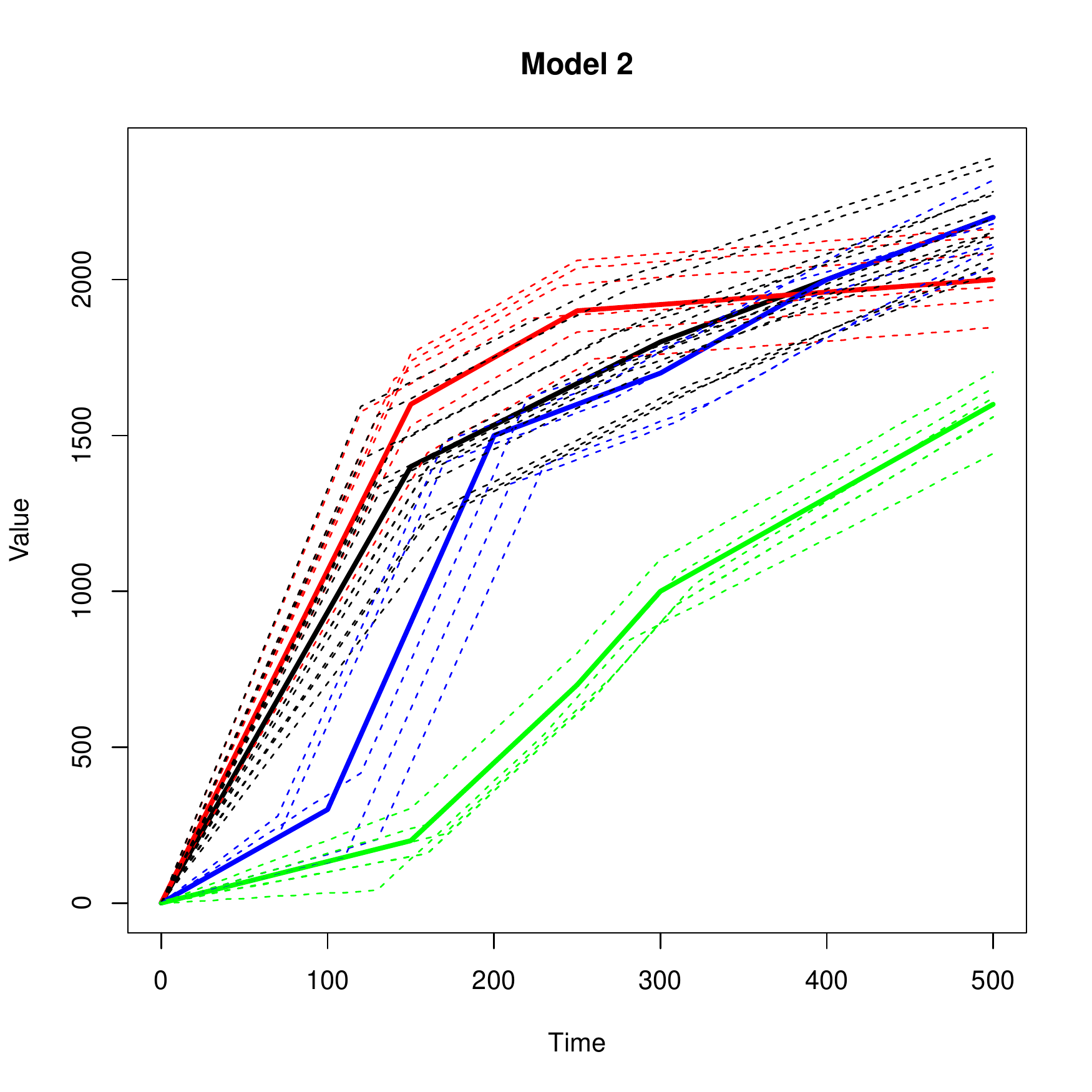} & \includegraphics[width = 7.5cm, scale=0.1]{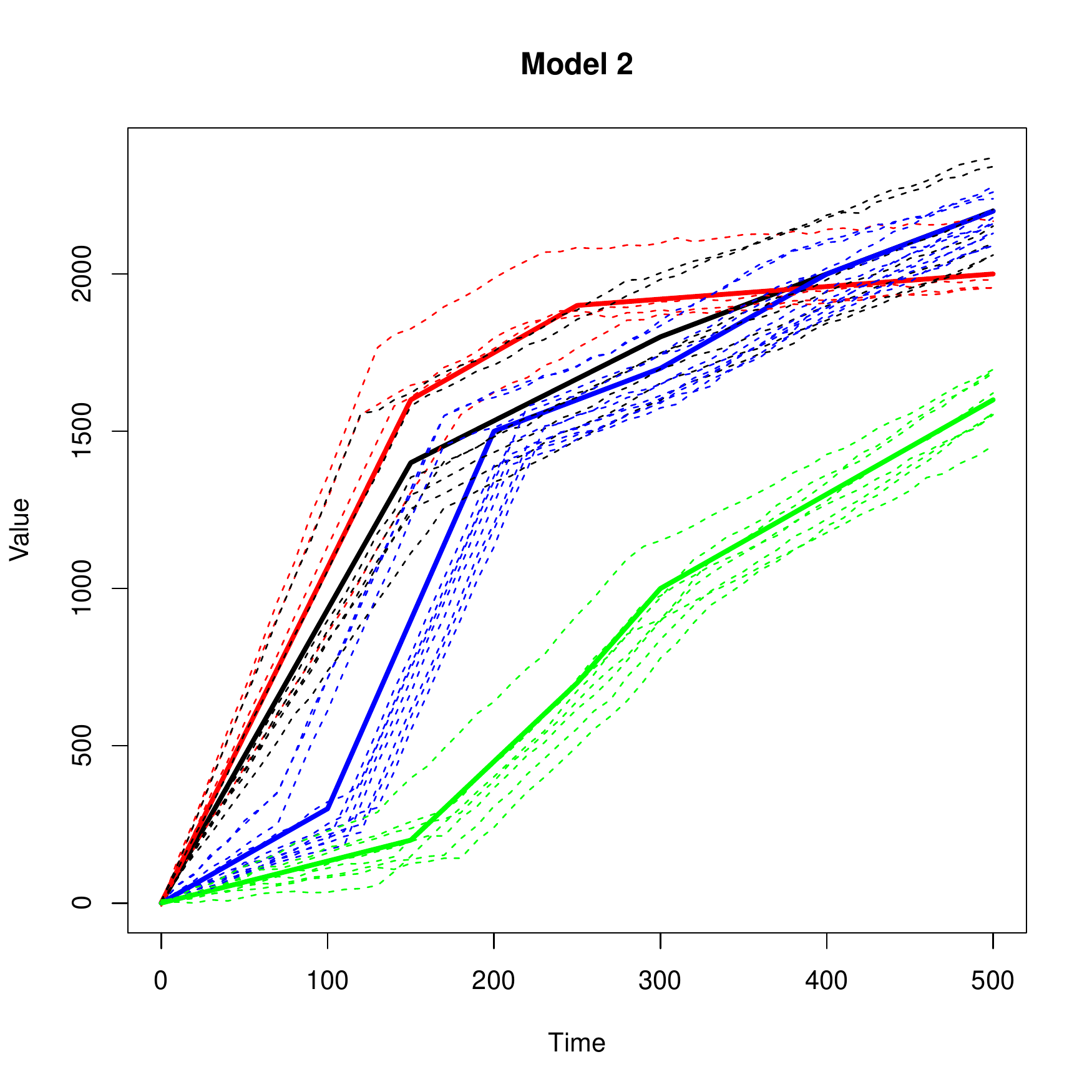}
\end{tabular}
\end{center}
\caption{\label{fig:figModel}
  Examples of observations generated from $\texttt{Model 1}$ (top) and $\texttt{Model 2}$ (bottom) for $\sigma=1$ (left) and $\sigma=5$ (right).
The curves belonging to Cluster 1 (resp. 2, 3, 4) are displayed in red (resp. black, blue and green). The solid lines display the representative curves
of each cluster $f_{t^z, \theta^z}$ and the dashed ones are some examples of the corresponding $\mathbf{Y}$.
}
\end{figure}

Figure~\ref{fig:figModel} displays some observations generated using the above simulation scheme for each model and for each $\sigma$.
We can see from this figure that the clustering problem associated to $\texttt{Model 1}$ seems to be the most difficult.
In $\texttt{Model 1}$, the clusters are indeed completely mixed whereas in $\texttt{Model 2}$ Cluster $\mathcal{C}_4$
is well separated from the others. Observe also that the data that is generated has the same behavior as the data 
coming from our motivating application: They are nondecreasing and piecewise linear constant with a small additive noise, see Figure \ref{fig1}.


\subsection{Statistical performance}
\label{subsec:numericalPerf}

Following the simulation scheme described in Section~\ref{subsec:simulScheme},
the performance of our procedure is assessed for each model, each $\sigma$ and is compared with
two different clustering methods: the $k$-means algorithm applied to the raw data $\mathbf{Y}$ and 
the FunFEM procedure described in \cite{bouveyron2015} and available in the \texttt{R} package \texttt{FunFEM}. The latter method is dedicated to the clustering of functional data and 
is based on a functional mixture model. All the methods are compared thanks to the Adjusted Rand Index (ARI) defined in \cite{Hubert1985}  which is often used for clustering validation. 
It is indeed a measure of agreement between two partitions. Note that the number of clusters $k$ in the $k$-means algorithm is chosen using the same strategy as the one that we considered
in our approach. As far as \texttt{FunFEM} is concerned, we used the default parameters.

For each model and for each $\sigma$ in $\{1,5\}$, we repeat independently 100 times the following steps:  
\begin{enumerate}
\item[(a)] We simulate a sample $\mathcal{D}_N = \{(\mathbf{Y}^{1},Z^{1}) \ldots (\mathbf{Y}^{N},Z^{N})\}$ of size $N = 100$ according to 
the scheme described in Section~\ref{subsec:simulScheme}; 
\item[(b)] We apply each method to $\mathcal{D}_N$;
\item[(c)] Based on the obtained clustering, we compute the ARI.
\end{enumerate}

\begin{figure}
\begin{center}
\begin{tabular}{cc}
\includegraphics[width = 7.5cm, scale=1]{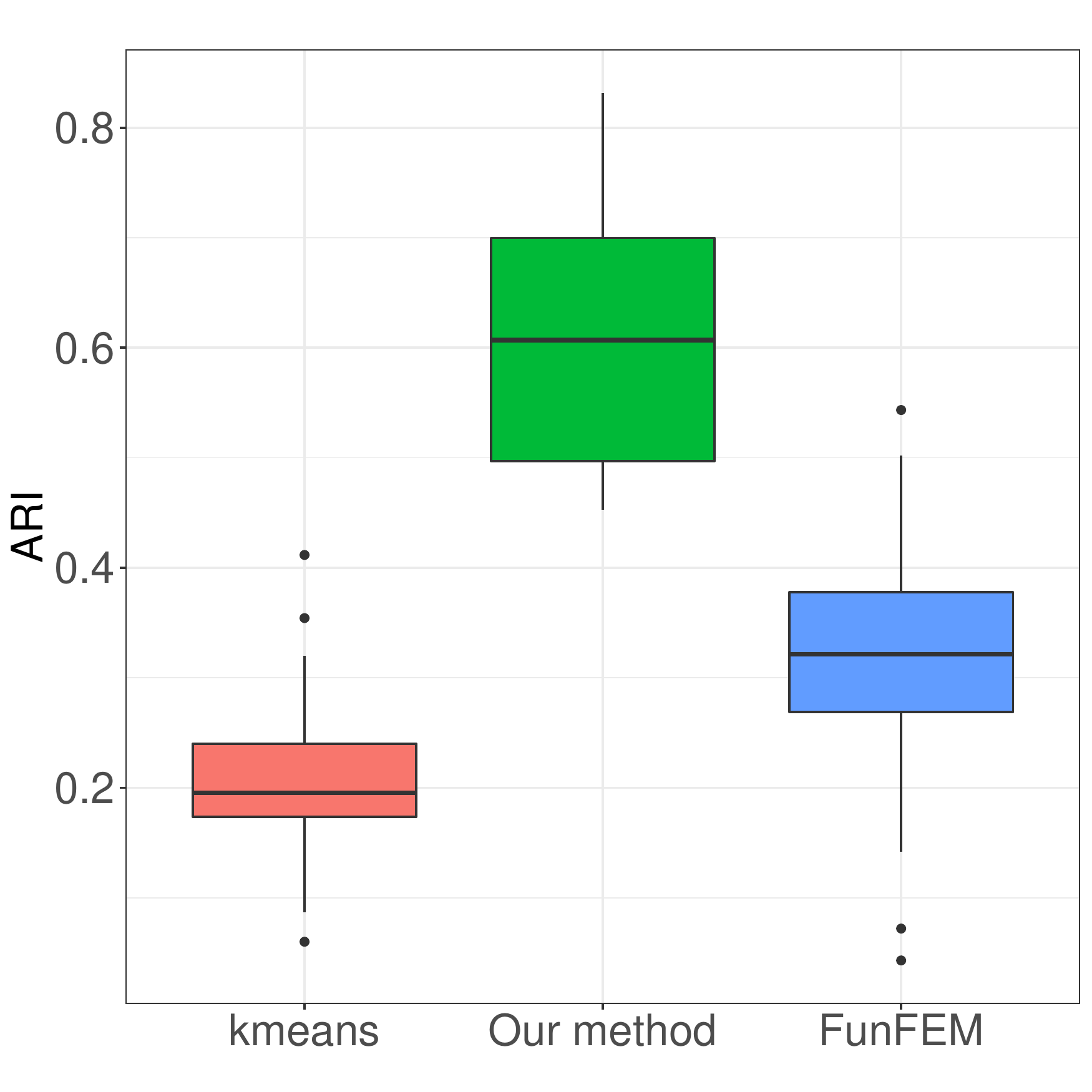}  & \includegraphics[width = 7.5cm, scale=0.1]{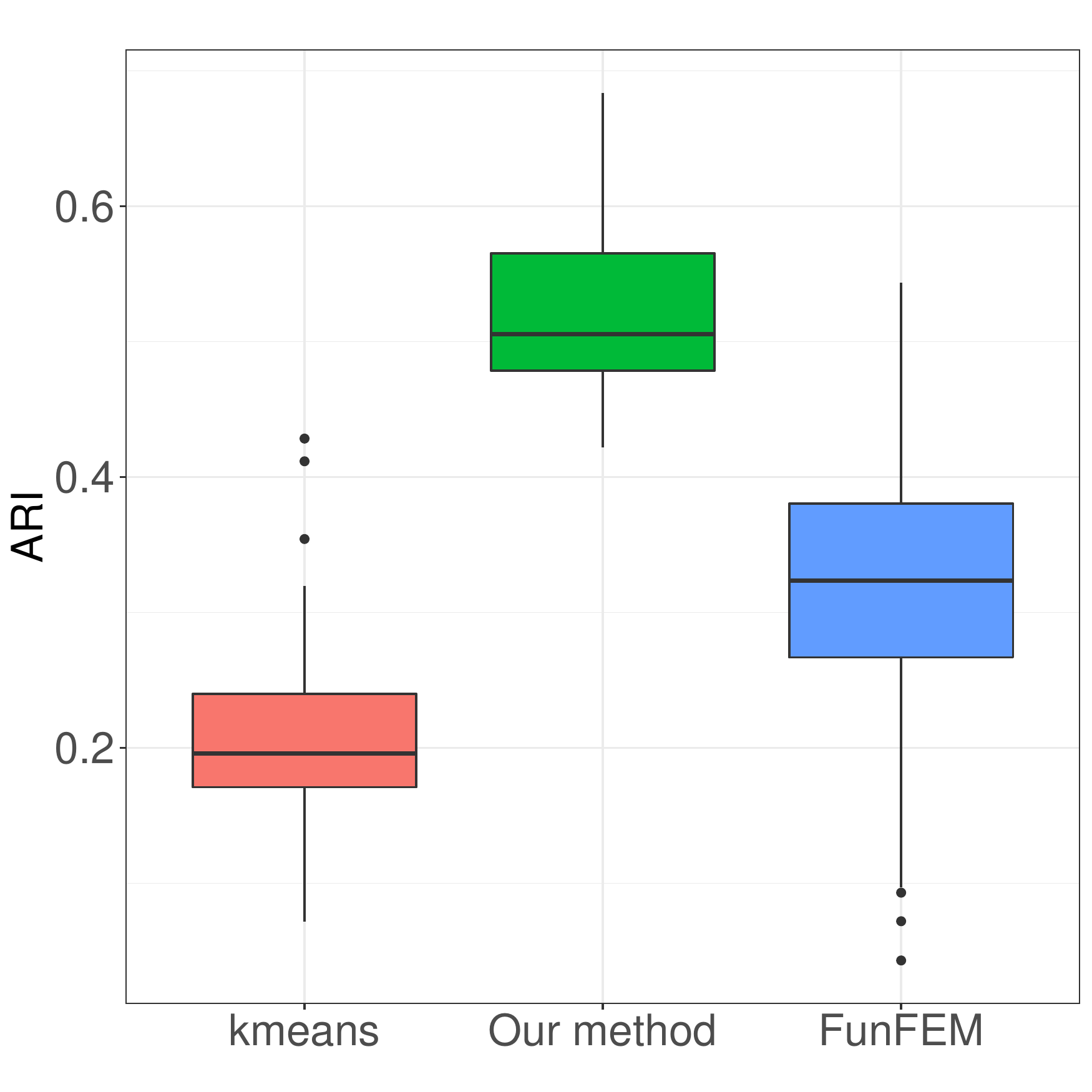} \\
\includegraphics[width = 7.5cm, scale=1]{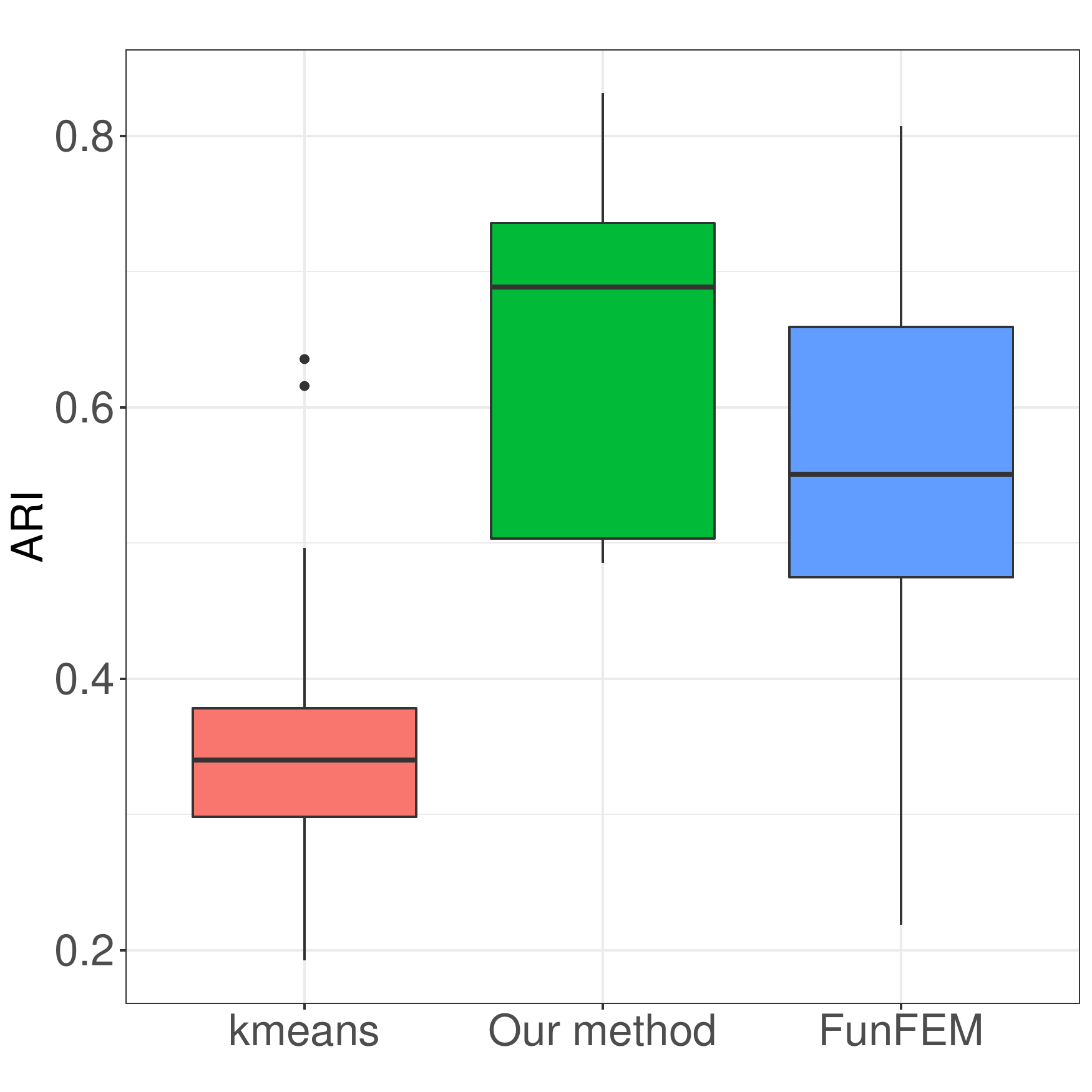}  & \includegraphics[width = 7.5cm, scale=0.1]{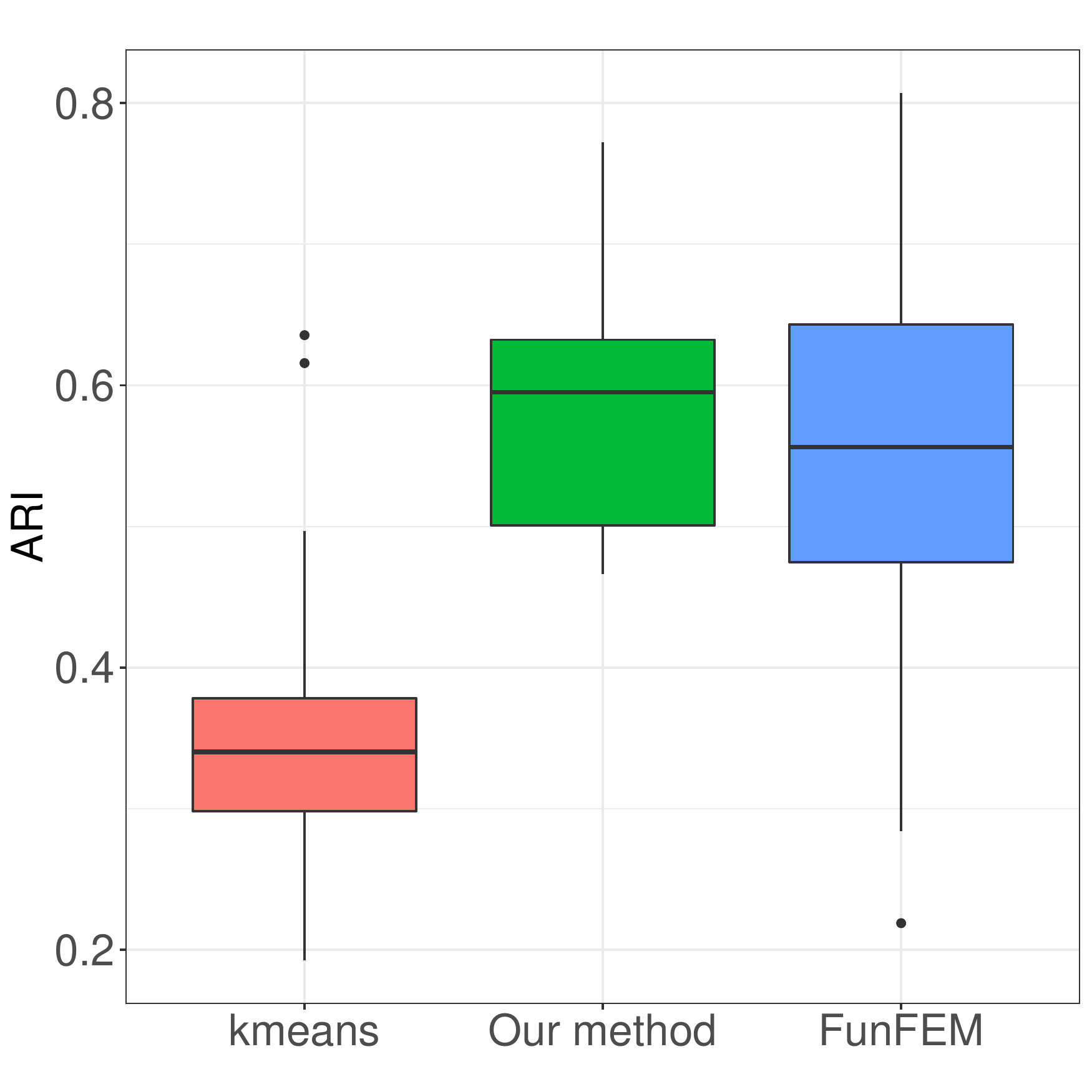}
\end{tabular}
\end{center}
\caption{\label{fig:figResult}
Boxplots of the ARI for $\texttt{Model 1}$ (top) and $\texttt{Model 2}$ (bottom) for $\sigma=1$ (left) and $\sigma=5$ (right).}
\end{figure}

The results are displayed in Figure~\ref{fig:figResult} with $K_{\textrm{max}}=10$.
We can see from this figure that our method outperforms the other ones in all cases except for Model 2 with $\sigma=5$ where the performance of our method is on a par with the one of FunFEM.
Note that applying the $k$-means to a relevant summary measure of $\mathbf{Y}$ significantly improves the clustering performance. Moreover, we observe that when $\sigma$ increases, the performance
of our approach is slightly altered since the change-points are more difficult to locate accurately, see Section \ref{sec:assessment}.

\subsection{Assessment of our change-point estimation procedure}
\label{sec:assessment}

We provide the following numerical experiments for assessing the change-point estimation stage of our method. We used the parameters associated to Cluster 3 of Model 1, see Table 
\ref{table:tableParameters1}. We repeat 100 times
\begin{enumerate}
\item[(a)] We simulate $\mathbf{Y}$ according to Equation (\ref{eq:eqSimY}) with $\sigma \in \{1,5\}$;
\item[(b)] We estimate the change-points according to the procedure described in the first stage of the first step in Section~\ref{sec:methodology}.
\end{enumerate}

\begin{figure}
\includegraphics[scale=0.4,trim={0 0 0 2cm},clip]{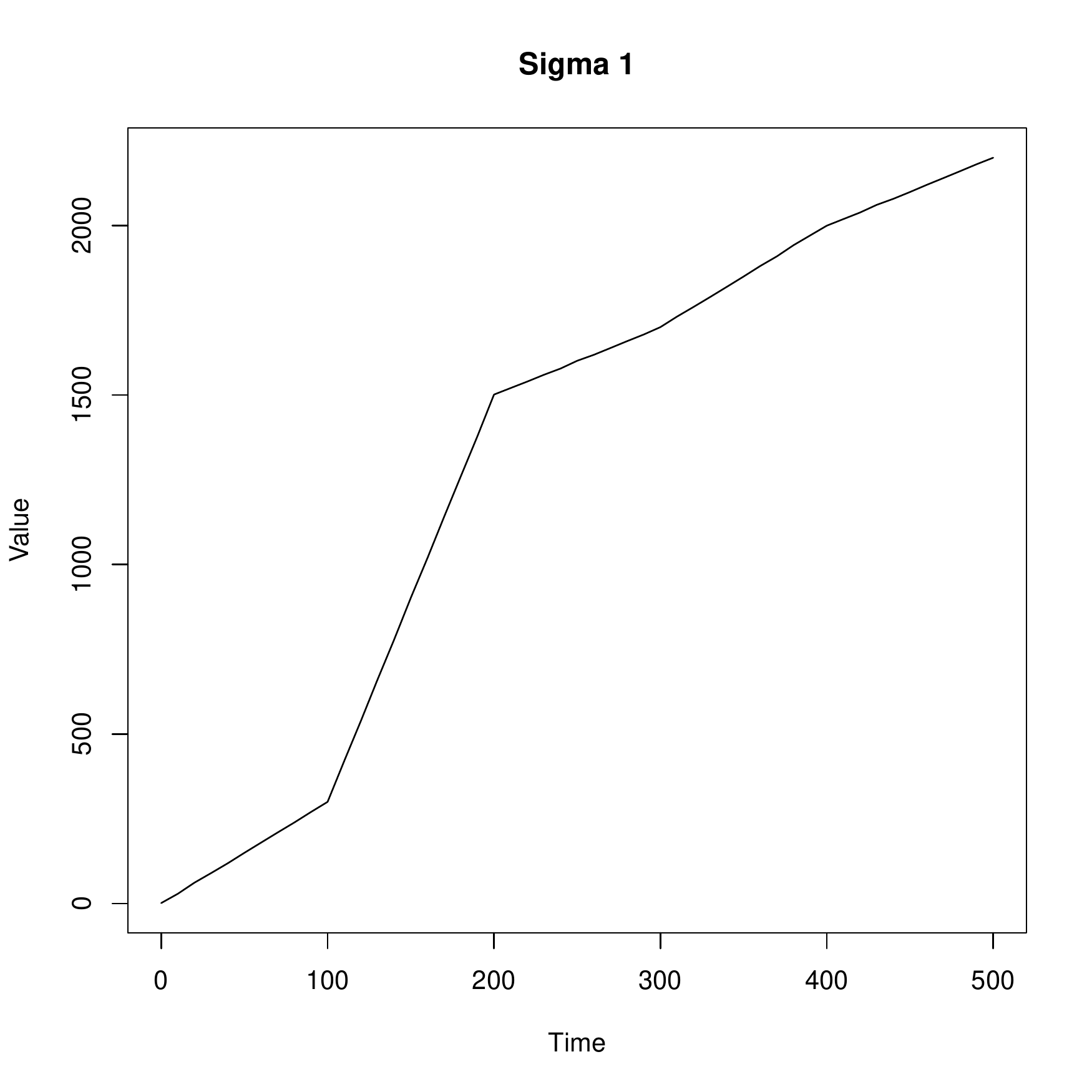}
\includegraphics[scale=0.4,trim={0 0 0 2cm},clip]{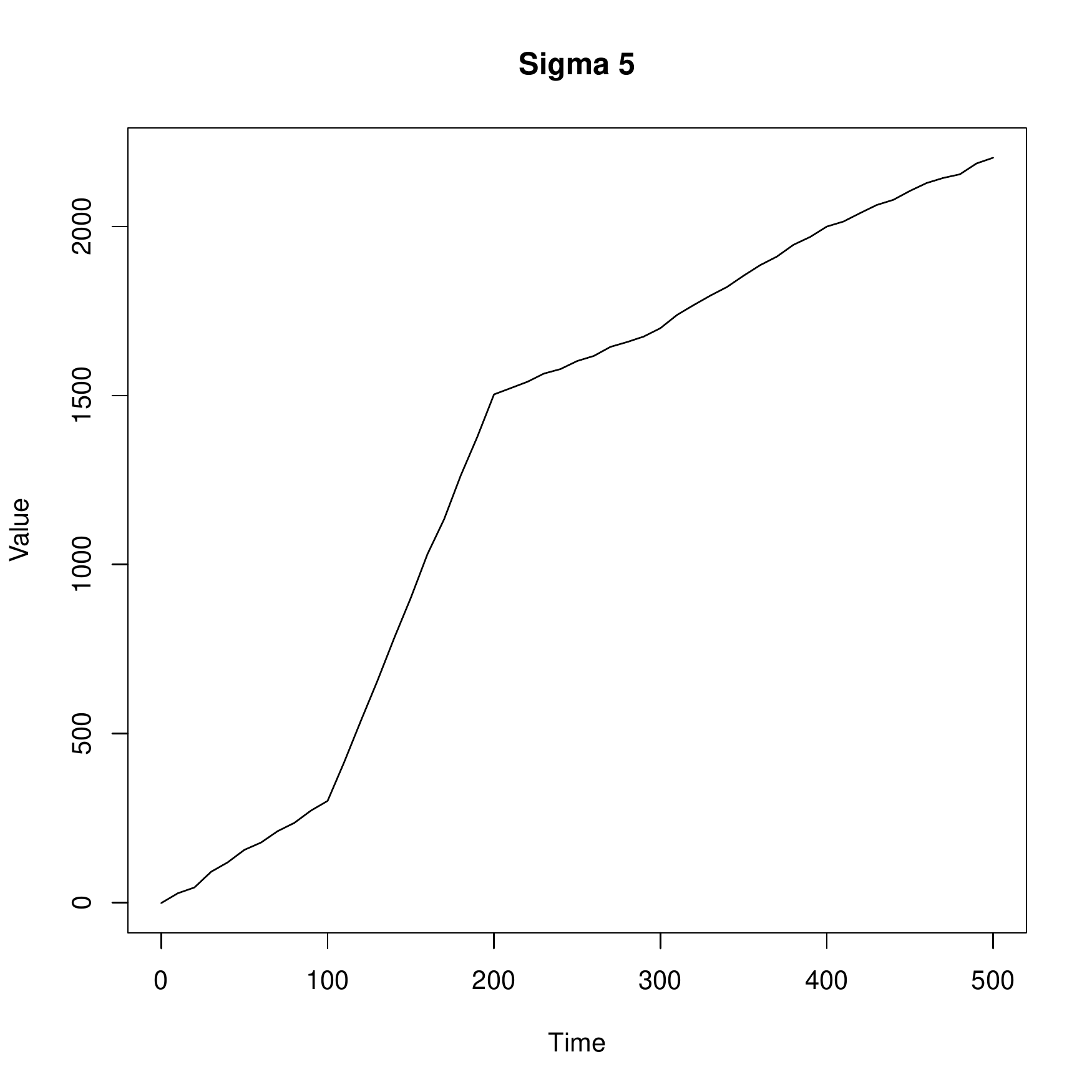}
\caption{\label{fig:extraj} Examples of $\mathbf{Y}$ belonging to Cluster 3 of Model 1 for $\sigma=1$ (left) and $\sigma=5$ (right).}
\end{figure}

Some examples of $\mathbf{Y}$ for the two values of $\sigma$ are displayed in Figure \ref{fig:extraj}. We can see from this figure that the change-points located at 300 and 400 are
more difficult to detect than the others. It is all the more true when $\sigma=5$.

\begin{figure}
\begin{center}
\begin{tabular}{cc}
\includegraphics[width = 7.5cm, scale=1]{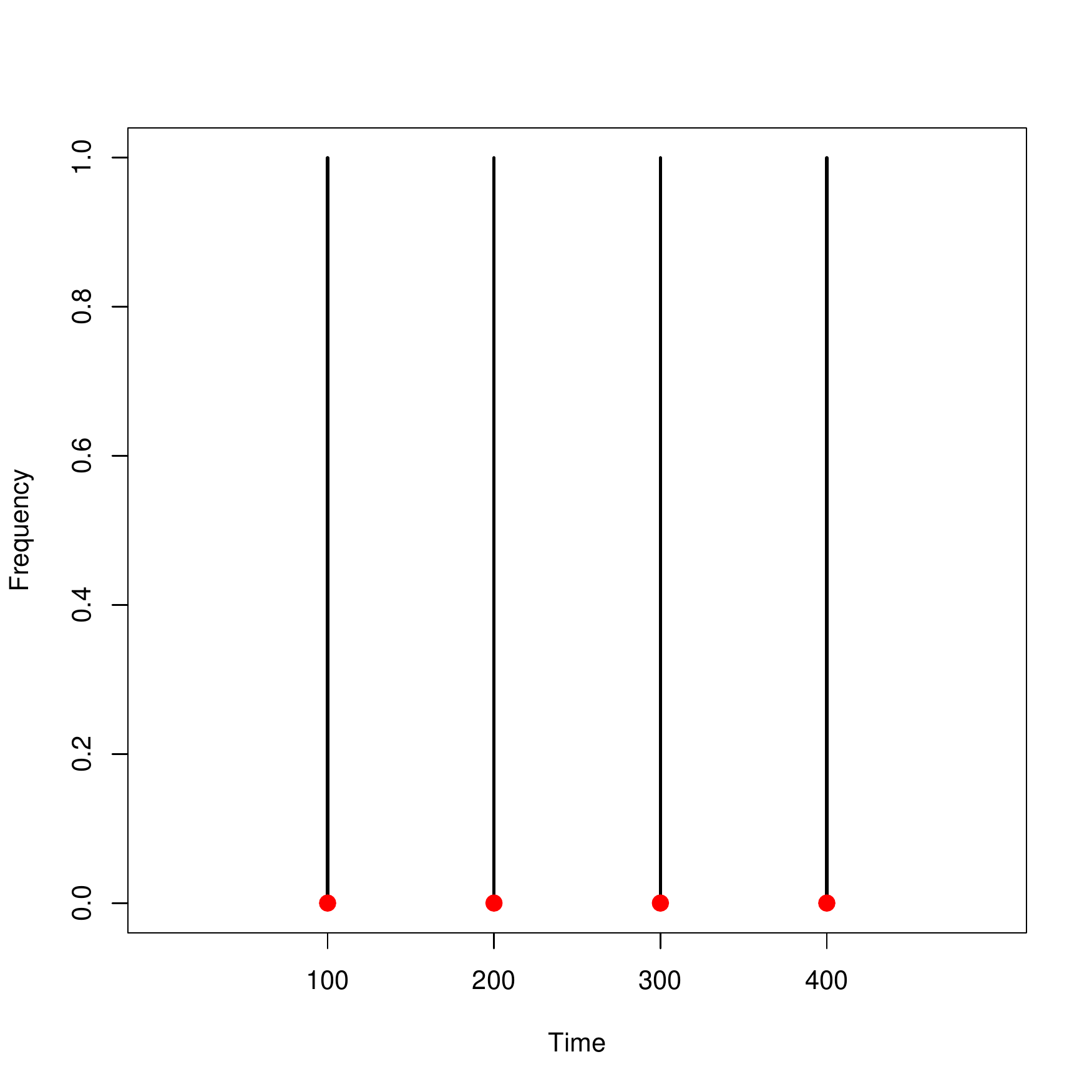}  & \includegraphics[width = 7.5cm, scale=0.1]{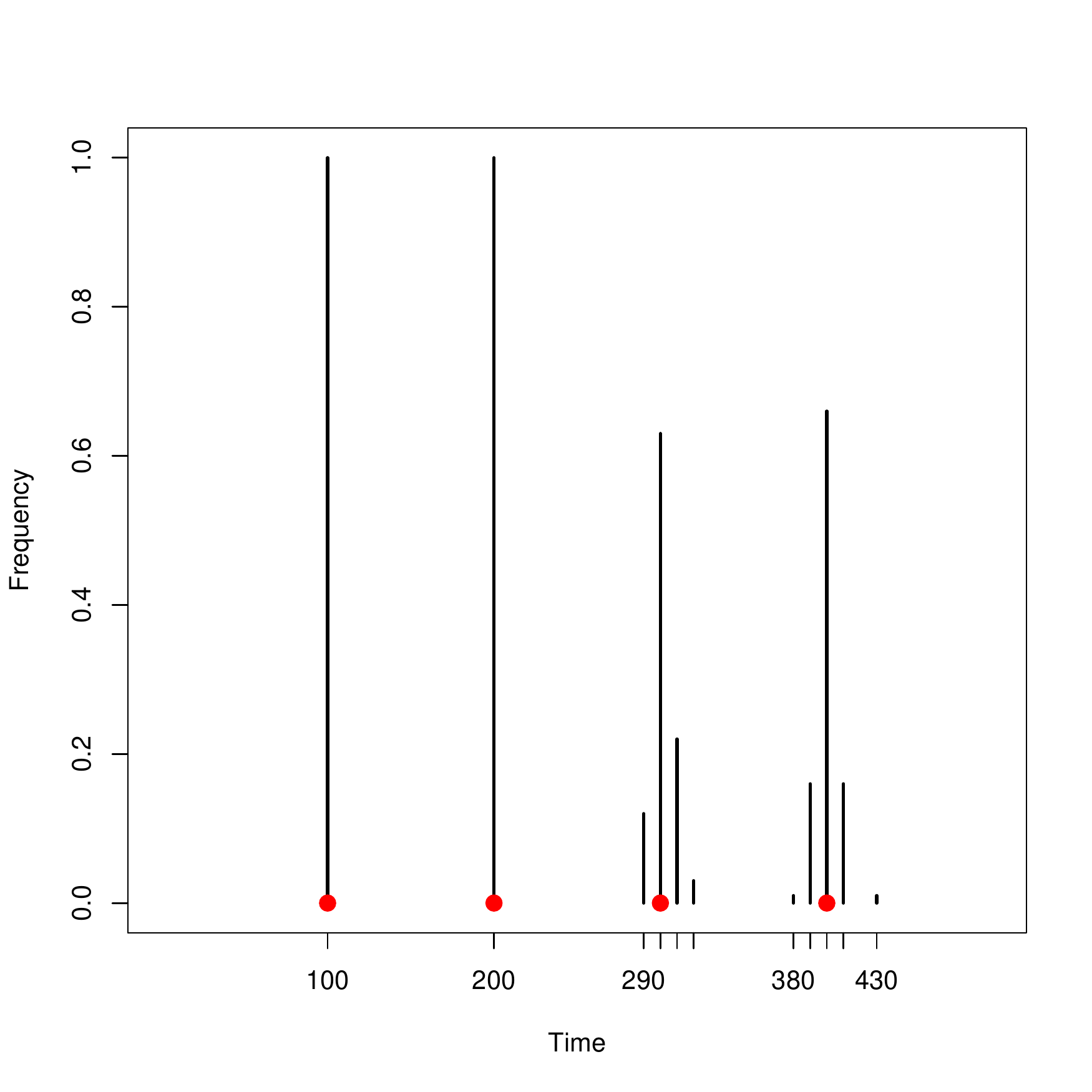} 
\end{tabular}
\end{center}
\caption{\label{fig:figChangePoint}
Change-point estimation frequencies for $\sigma=1$ (left) and $\sigma=5$ (right). The true change-point positions are denoted with red plain circles.}
\end{figure}

Figure~\ref{fig:figChangePoint} displays the frequency of the number of times where each position 
 has been estimated as a change-point. We can see that the change-points are all retrieved and that no spurious change-points are provided when
$\sigma=1$.  In the case where $\sigma=5$, although the positions of the true change-points are retrieved most of the time,
some additional spurious change-points are also selected with a very low frequency.

\section{Application}\label{sec:appli}

In this section, we apply the methodology described in Section \ref{sec:methodology} to milking kinetics of  dairy goats coming from 
the experimental herd of the research unit Systemic Modelling Applied to Ruminants (Paris, France).

\subsection{Data description}  The data set contains 100470 
milking kinetics of goats of two different breeds:  ``Alpine'' and ``Saanen''. All these kinetics are morning milking kinetics and several kinetics are available for each goat.
The kinetics can also be separated according to parity which corresponds to the lactation rank \textit{i.e.} to the number of times a goat has given birth and started a new lactation.
In the considered dataset, there are in particular 276 (resp. 191) goats for which we have their milking kinetics for Parity 1 (resp. 2). 

\subsection{Kinetics clustering}
\begin{figure}
\includegraphics[scale=1]{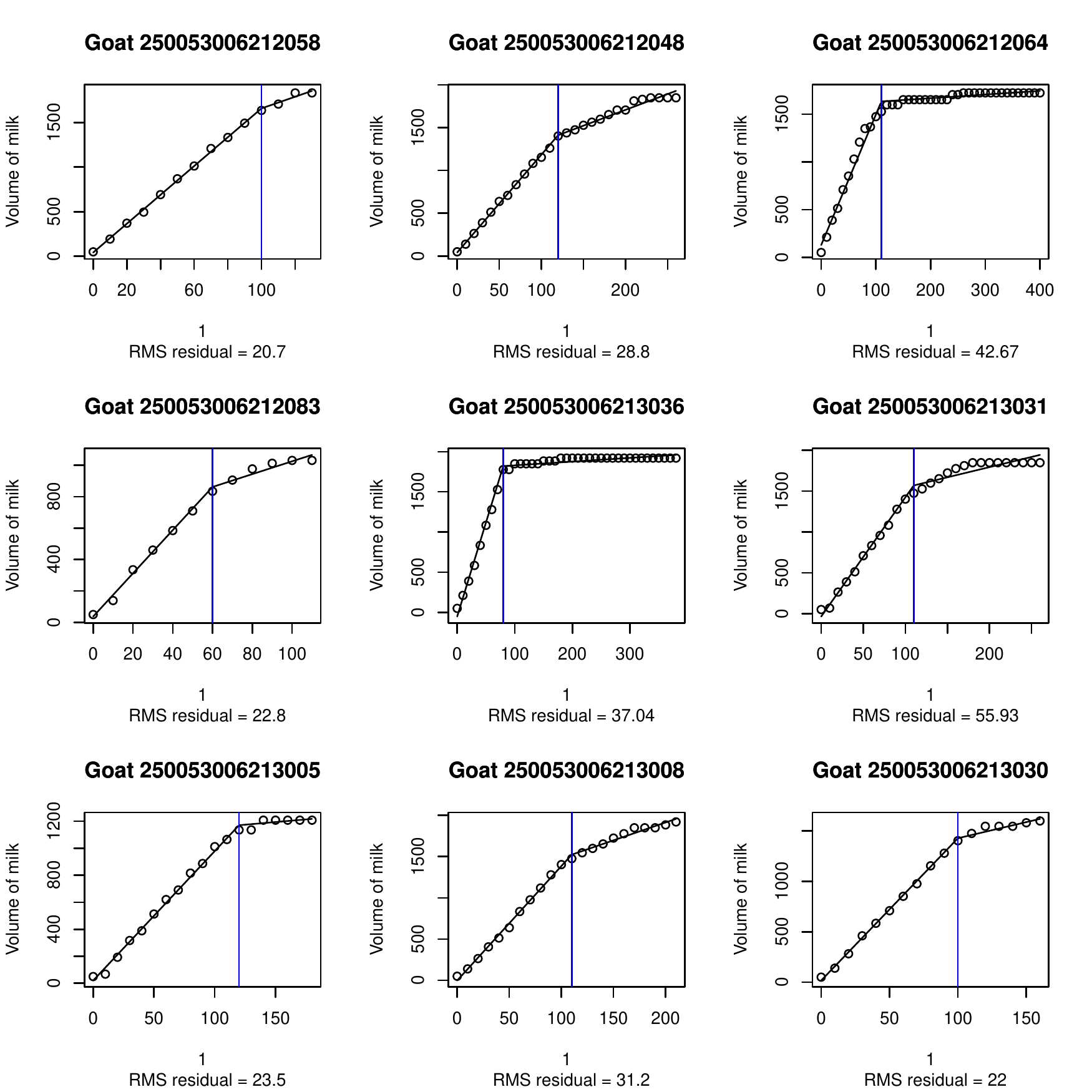}
\caption{Some examples of milking kinetics belonging to Cluster 1. The data are displayed with 'o', the straight lines correspond to the piecewise 
linear fit obtained thanks to our method and the vertical line corresponds to the position of the change-point.\label{fig:cluster1}}
\end{figure}

\begin{figure}
\includegraphics[scale=1]{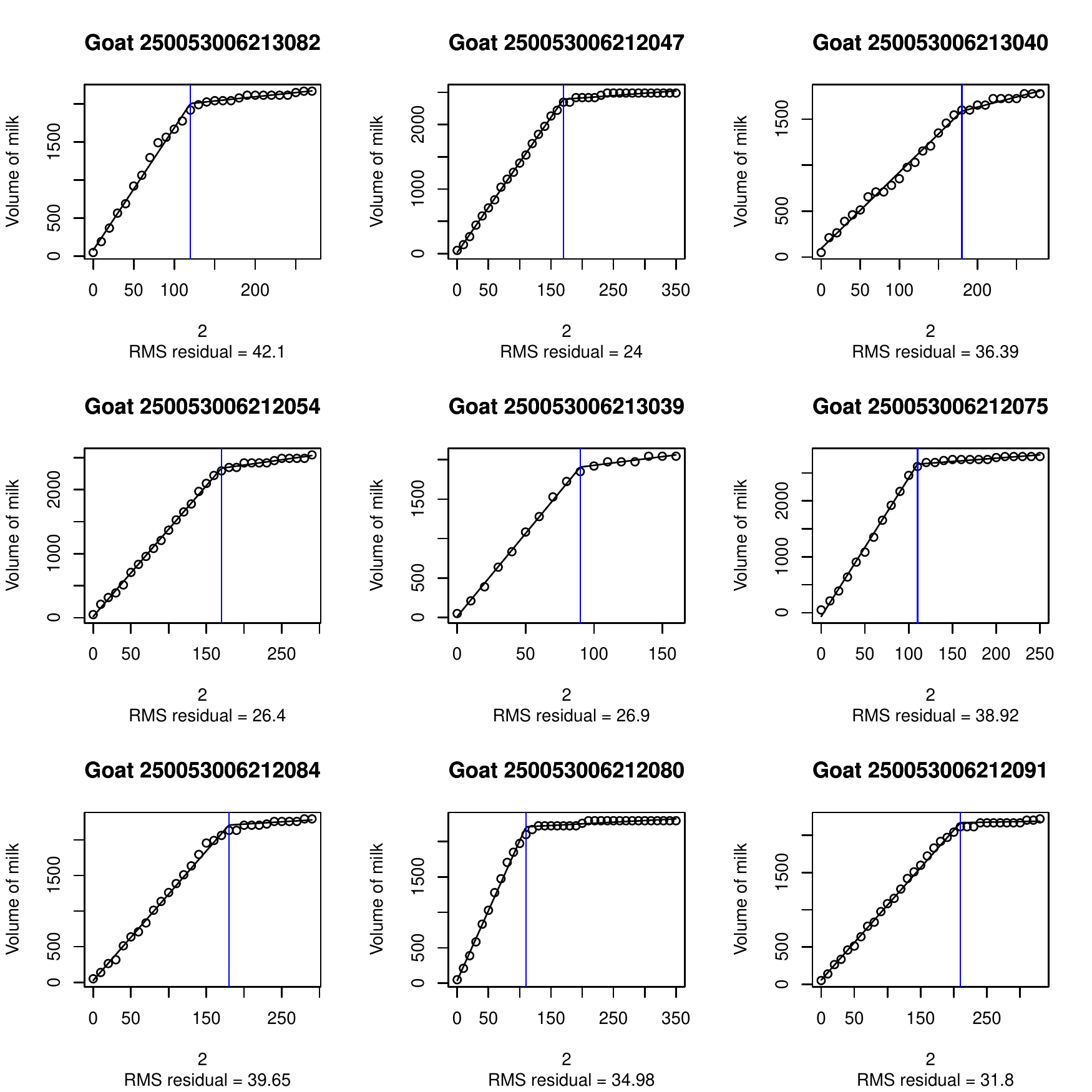}
\caption{Some examples of milking kinetics belonging to Cluster 2. The data are displayed with 'o', the straight lines correspond to the piecewise 
linear fit obtained thanks to our method and the vertical line corresponds to the position of the change-point.\label{fig:cluster2}}
\end{figure}

\begin{figure}
\includegraphics[scale=1]{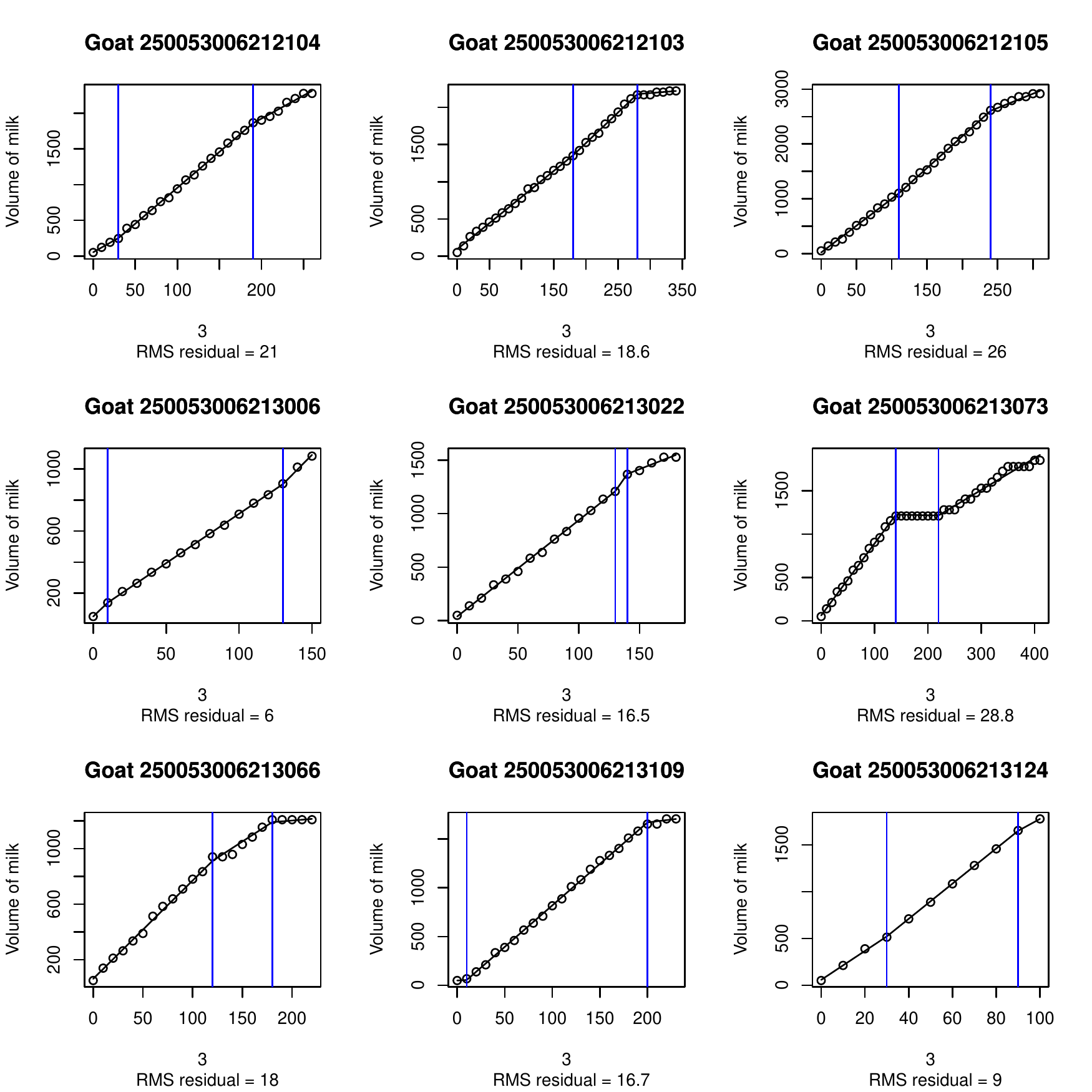}
\caption{Some examples of milking kinetics belonging to Cluster 3. The data are displayed with 'o', the straight lines correspond to the piecewise 
linear fit obtained thanks to our method and the vertical line corresponds to the position of the change-points.\label{fig:cluster3}}
\end{figure}

\begin{figure}
\includegraphics[scale=0.5]{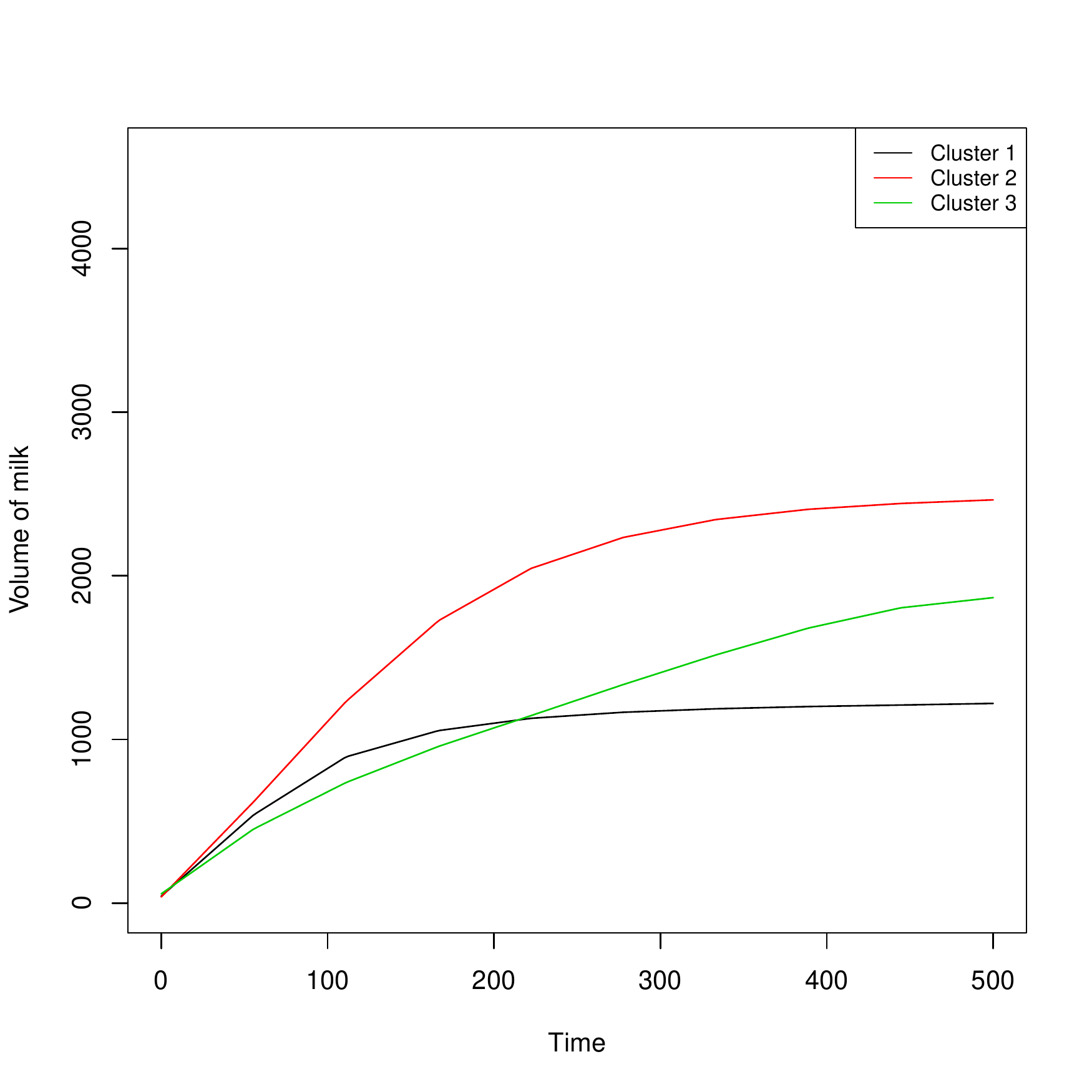}
\caption{Kinetics average obtained within each of the three clusters.\label{fig:mean_cluster}}
\end{figure}

First, note that based on the shapes of the milking kinetics of this data set, the parameter  $K_{\textrm{max}}$
defined in the first stage of the first step in Section \ref{sec:methodology} was set to 2. We obtained three clusters containing 57498, 36757 and 6215 kinetics, respectively. Some examples of kinetics belonging to Clusters 1, 2 and 3 are
displayed in Figures \ref{fig:cluster1}, \ref{fig:cluster2} and \ref{fig:cluster3}, respectively. The average of the kinetics estimations obtained within each cluster is displayed in Figure \ref{fig:mean_cluster}. We can observe that the three clusters can be distinguished in terms of quantity of milk production: Cluster 1 has the lowest production, Cluster 3 the highest and Cluster 2 is 
between them.

\begin{figure}
\includegraphics[scale=0.4]{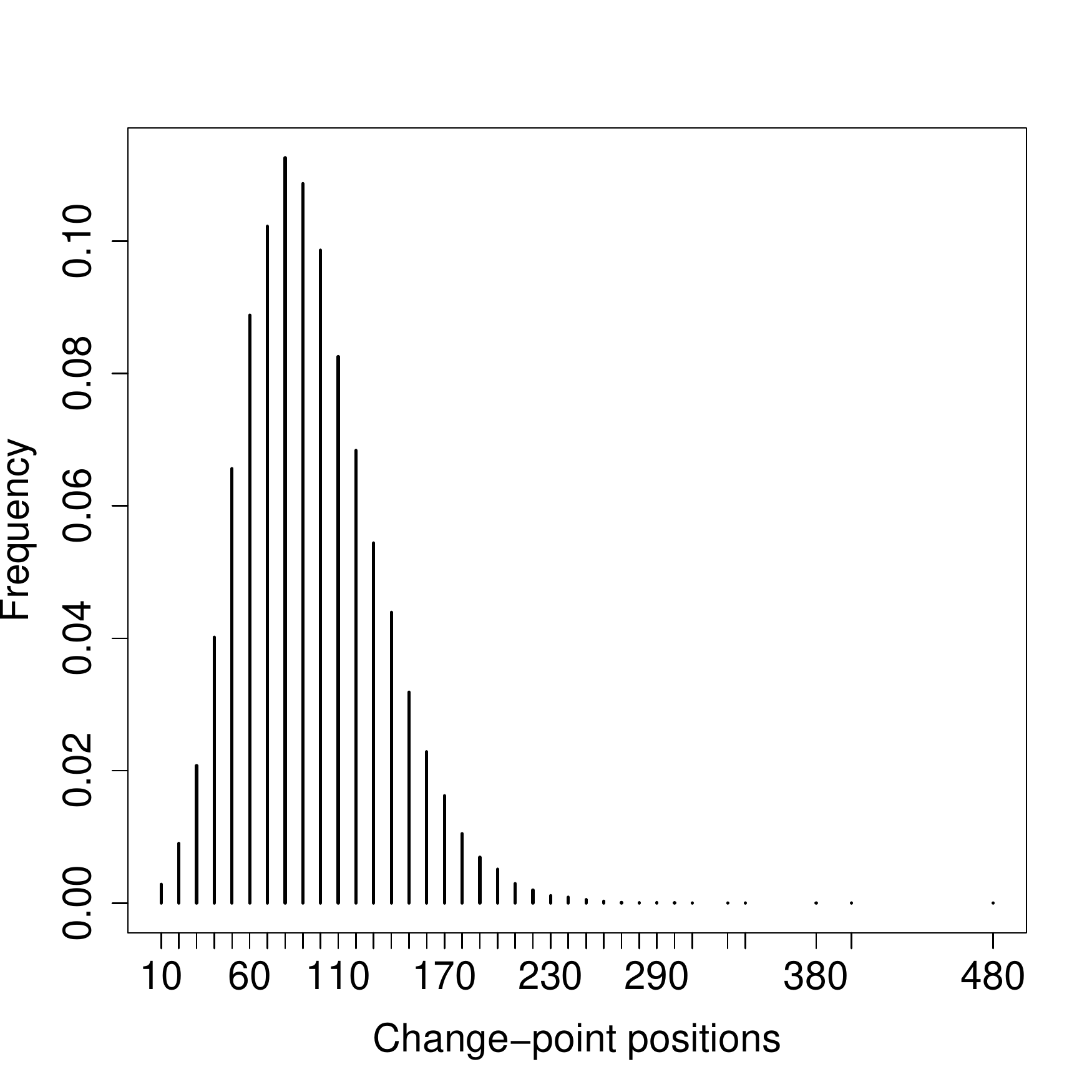}
\includegraphics[scale=0.4]{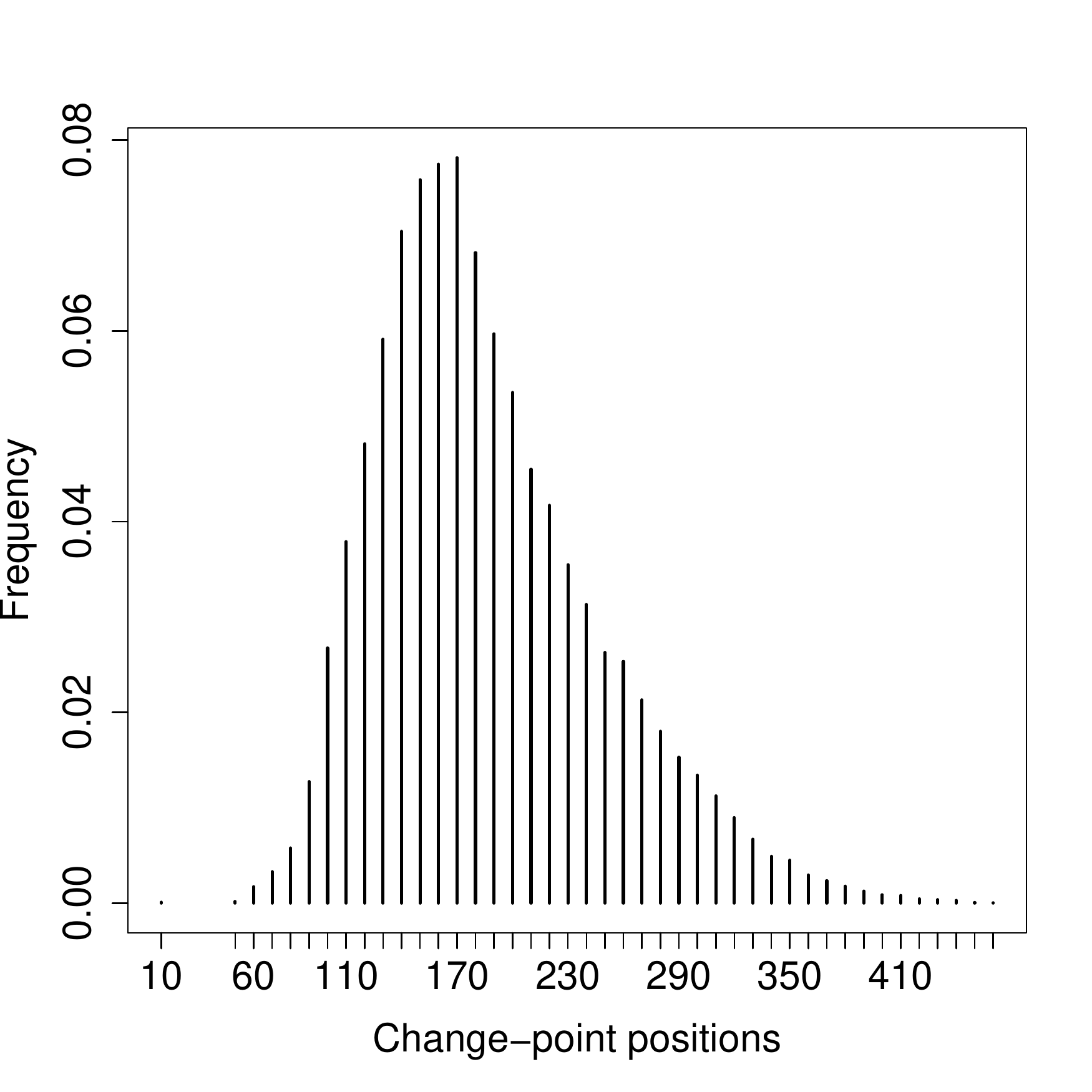}
\caption{Histograms of the change-point positions for Cluster 1 (left) and Cluster 2 (right).\label{fig:histogram}}
\end{figure}

Another difference between the three clusters is the number and the positions of changes. The number of changes in the kinetics of Cluster 1 and 2 is mainly one contrary to Cluster 3 where this number is always equal to two. Figure \ref{fig:histogram} displays the histogram of the change-point positions for Clusters 1 and 2.  We can observe that the change-point having the highest frequency
is not located at the same position for these two clusters.
Interestingly, our methodology was able to distinguish these two clusters thanks to the change-point
position which illustrates the potential of our methodology to extract synthetic traits from raw data.

In practice, such a clustering may be very useful in the precision farming context
to refine selection criteria for breeding programs, to simplify milking workload or to control udder health.
Thanks to the clustering results, we should be able to define a milking profile for each goat.
Moreover, we propose in the next section to characterize dairy goats belonging to a given parity.

\subsection{Parity characterization}

\begin{figure}
\includegraphics[scale=0.35]{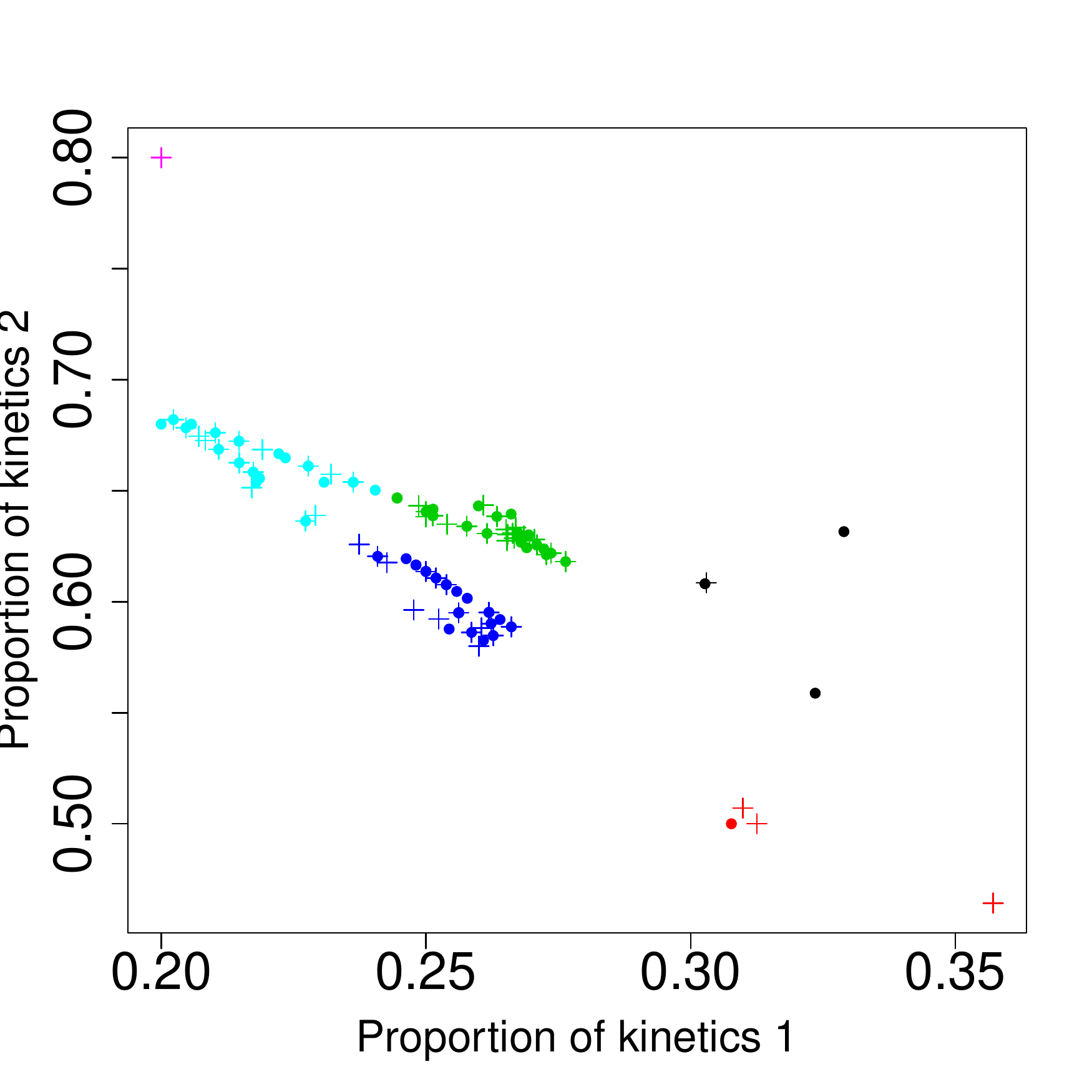}
\includegraphics[scale=0.35]{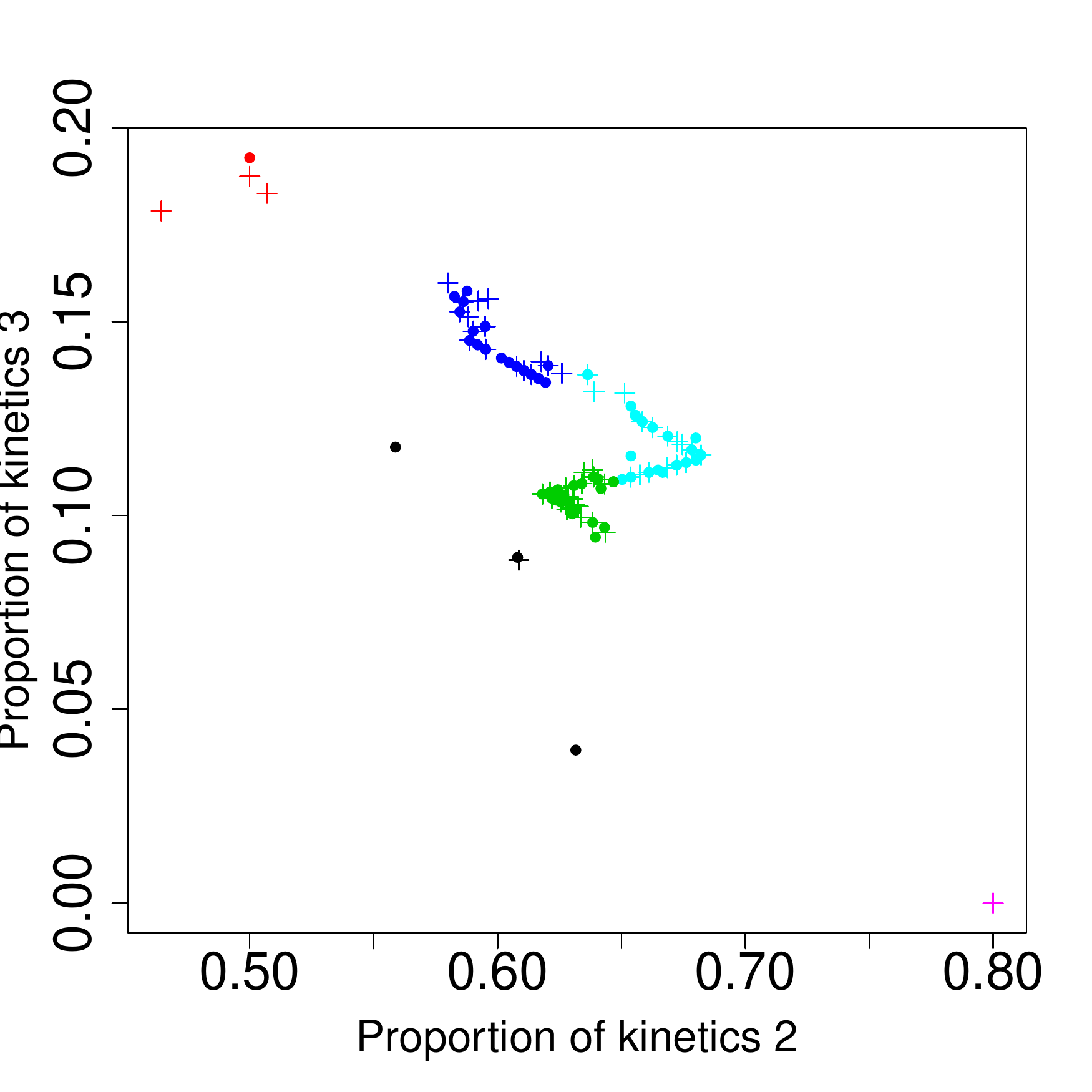}\\
\includegraphics[scale=0.35]{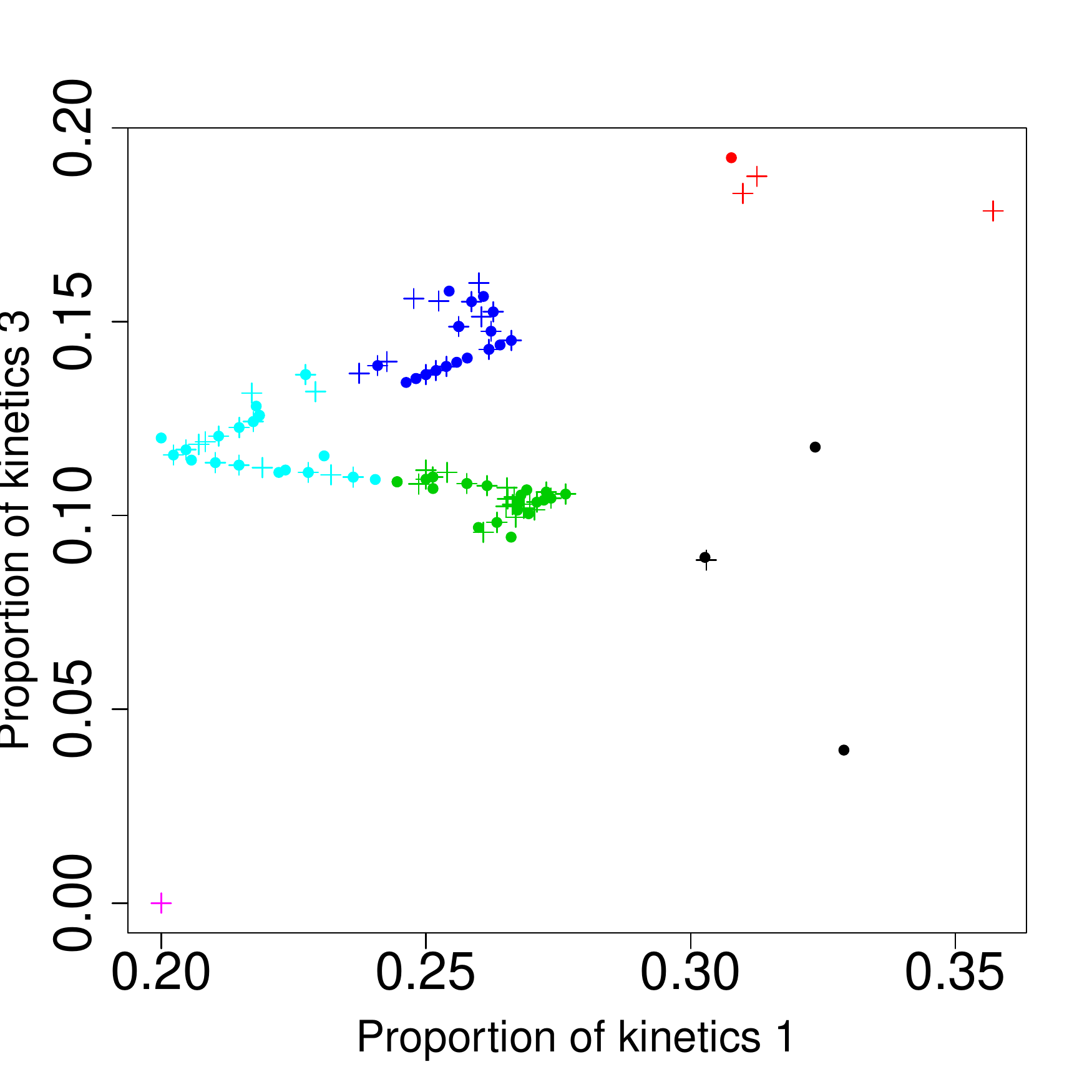}
\caption{Clustering obtained for goats in Parity 1 (6 clusters) displayed on the plane having for axes the proportion of kinetics belonging to Clusters 1 and 2 (top left), 2 and 3 (top right), 1 and 3 (bottom). The Saanen (resp. Alpine) goats are displayed with '$\bullet$' (resp. '+'). \label{fig:parite1_1fonc_2}}
\end{figure}

\begin{figure}
\includegraphics[scale=0.35]{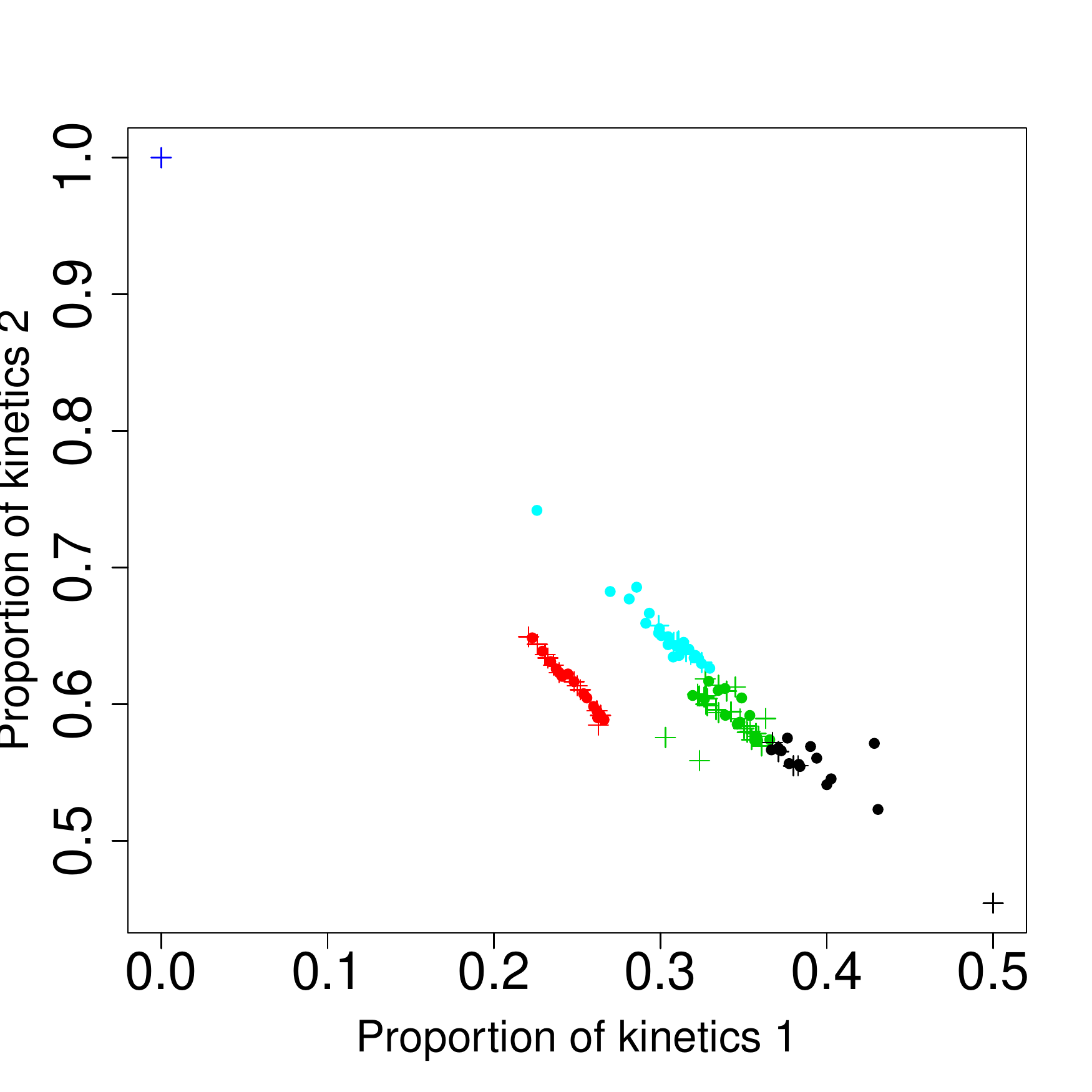}
\includegraphics[scale=0.35]{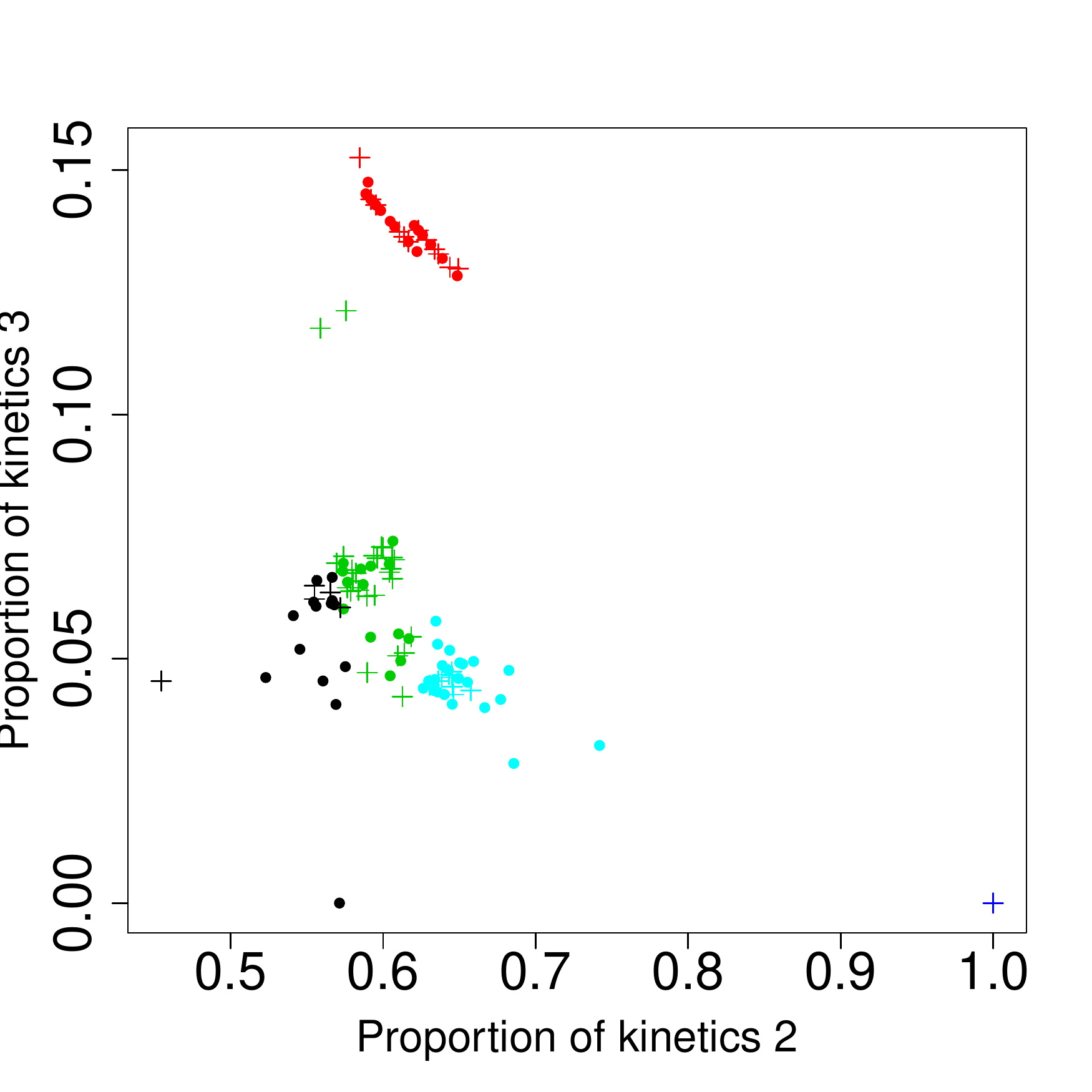}\\
\includegraphics[scale=0.35]{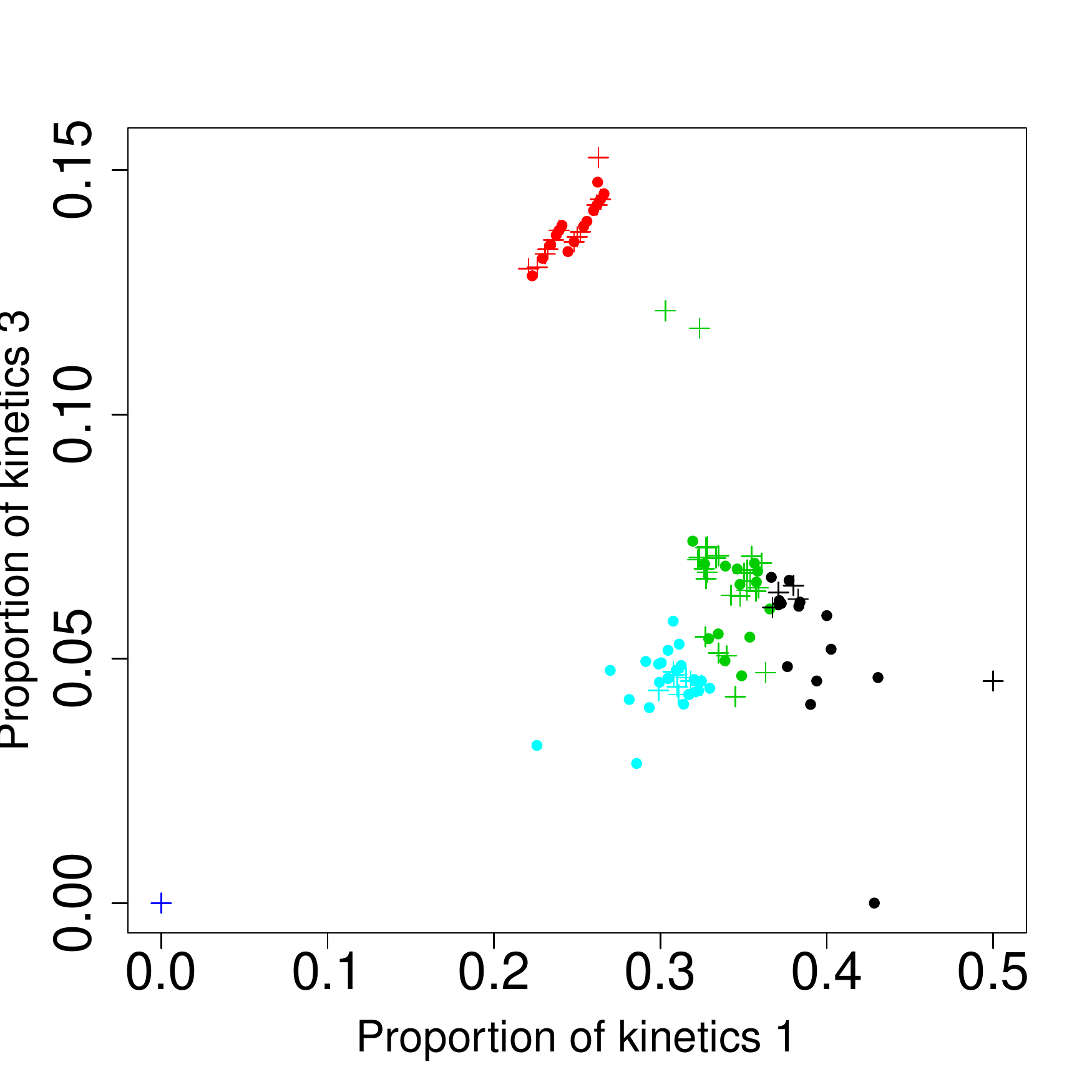}
\caption{Clustering obtained for goats in Parity 2 (5 clusters) displayed on the plane having for axes the proportion of kinetics belonging to Clusters 1 and 2 (top left), 2 and 3 (top right), 1 and 3 (bottom). The Saanen (resp. Alpine) goats are displayed with '$\bullet$' (resp. '+').\label{fig:parite2_1fonc_2}}
\end{figure}

In order to go further into this analysis, we tried to characterize the parities 1 and 2 in terms of the proportion of kinetics of type 1, 2 or 3 according to the clustering previously obtained.
We thus created for each goat belonging to a given parity a vector of proportions corresponding to its belonging frequency to each Cluster 1, 2 or 3. 
For each parity, the goats are clustered using the $k$-means algorithm applied to the vectors of proportions. The results are displayed in Figures \ref{fig:parite1_1fonc_2} and \ref{fig:parite2_1fonc_2} for Parities 1 and 2, respectively. The number of groups is selected using the method described in Section \ref{subsec:second_step}: We found 6 (resp. 5) groups for Parity 1 (resp. 2).  We can notice that there is one goat which produces
a large quantity of milk compared to the others for both parities. In Parity 1, 80\% of its milking kinetics
belong to Cluster 2 and only 20\% to Cluster 1. In Parity 2, 100\% of its milking kinetics belong to Cluster 2.

 We also observe from Figures \ref{fig:parite1_1fonc_2} and \ref{fig:parite2_1fonc_2} that in both parities, the belonging frequency of the milking kinetics to Cluster 2 is between 50\% and 70\%.
In Parity 2, there is one group (in red) for which the proportion of milking kinetics belonging to Cluster 2 is very high (around 65\%) and the
proportions of milking kinetics belonging to Cluster 1 and Cluster 3 are very low (around 25\% and 13\%, respectively). For the other groups the
proportions of milking kinetics belonging to Cluster 1 are higher. In Parity 1, the behavior is a little bit different in the sense that the majority of goats
have a high proportion of milking kinetics belonging to Cluster 2 (around 65\%)  and a low proportion of milking kinetics belonging to Cluster 1 (around 25\%).
Such results may be interesting in the context of precision breeding since they could help to forecast the production of milk at the different parities.

Further analysis should be perfomed in the future to study how evolve the cluster belonging along the lactation course lasting around 150 days in goats. The daily milk yield of a goat for a given parity follows indeed a typical triphasic shape (respectively increasing, plateau and decreasing phase), each daily milk yield being the sum of the total milk produced during each milking (respectively morning and afternoon milking). Being able to link a particular shape at the milking kinetics scale with one at the lactation scale could open perspectives to better characterize individual goats and thus propose options for individual milking management.

\bibliographystyle{chicago}
\bibliography{}

\end{document}